\documentclass[twocolumn]{aastex63}

\usepackage{natbib}
\usepackage{amsmath}
\usepackage{multirow}
\usepackage[normalem]{ulem}
\usepackage{xcolor}
\usepackage{xspace}
\usepackage{enumitem}   
\usepackage{graphicx}
\usepackage{wrapfig}
\usepackage{lineno}

\newcommand{\Rearth}{$R_\oplus$}

\newcommand{\Msun}{$M_\odot$}
\newcommand{\wotan}{\texttt{W\={o}tan}\xspace}
\newcommand{\epos}{\texttt{epos}\xspace}
\newcommand{\eleanor}{\texttt{eleanor}\xspace}
\newcommand{\gaia}{\texttt{Gaia}\xspace}
\newcommand{\pterodactyls}{\texttt{pterodactyls}\xspace}
\newcommand{\tls}{\texttt{TLS}\xspace}

\newcommand{\kepler}{\emph{Kepler}\xspace}
\newcommand{\spitzer}{\emph{Spitzer}\xspace}
\newcommand{\ktwo}{\emph{K2}\xspace}
\newcommand{\rotrate}{$R_{\text{rot}}$}
\newcommand{\triceratops}{\texttt{triceratops}\xspace}
\newcommand{\exotic}{\texttt{EXOTIC}\xspace}
\newcommand{\rprs}{$\frac{R_\text{p}}{R_\star}$}

\accepted{for publication in The Astronomical Journal; May 31st, 2022}
\definecolor{tropicalrainforest}{rgb}{0.0, 0.46, 0.37}
\definecolor{plum}{HTML}{88498f}

\shorttitle{\pterodactyls}
\shortauthors{Fernandes et al. 2022}

\begin{document}
\title{\pterodactyls: A Tool to Uniformly Search and Vet for Young Transiting Planets In TESS Primary Mission Photometry}

\correspondingauthor{Rachel B. Fernandes}
\email{rachelfernandes@email.arizona.edu}

\author[0000-0002-3853-7327]{Rachel B. Fernandes}
\affil{Lunar and Planetary Laboratory, The University of Arizona, Tucson, AZ 85721, USA}
\affil{Alien Earths Team, NASA Nexus for Exoplanet System Science, USA}

\author[0000-0002-1078-9493]{Gijs D. Mulders}
\affil{Facultad de Ingenier\'{i}a y Ciencias, Universidad Adolfo Ib\'{a}\~{n}ez, Av.\ Diagonal las Torres 2640, Pe\~{n}alol\'{e}n, Santiago, Chile}
\affil{Millennium Institute for Astrophysics, Chile}
\affil{Alien Earths Team, NASA Nexus for Exoplanet System Science, USA}

\author[0000-0001-7962-1683]{Ilaria Pascucci}
\affil{Lunar and Planetary Laboratory, The University of Arizona, Tucson, AZ 85721, USA}
\affil{Alien Earths Team, NASA Nexus for Exoplanet System Science, USA}

\author[0000-0003-4500-8850]{Galen J. Bergsten}
\affil{Lunar and Planetary Laboratory, The University of Arizona, Tucson, AZ 85721, USA}
\affil{Alien Earths Team, NASA Nexus for Exoplanet System Science, USA}

\author[0000-0003-3071-8358]{Tommi T. Koskinen}
\affil{Lunar and Planetary Laboratory, The University of Arizona, Tucson, AZ 85721, USA}
\affil{Alien Earths Team, NASA Nexus for Exoplanet System Science, USA}

\author[0000-0003-3702-0382]{Kevin K. Hardegree-Ullman}
\affil{Steward Observatory, The University of Arizona, Tucson, AZ 85721, USA}
\affil{Alien Earths Team, NASA Nexus for Exoplanet System Science, USA}

\author[0000-0002-5785-9073]{Kyle A. Pearson}
\affil{Jet Propulsion Laboratory, California Institute of Technology, 4800 Oak Grove Drive, Pasadena, CA 91109, USA}

\author[0000-0002-8965-3969]{Steven Giacalone}
\affil{Department of Astronomy, University of California Berkeley, Berkeley, CA 94720-3411, USA}

\author[0000-0003-1848-2063]{Jon Zink}
\affil{California Institute of Technology, Pasadena, CA, USA}

\author[0000-0002-5741-3047]{David R. Ciardi}
\affil{Caltech/IPAC-NASA Exoplanet Science Institute Pasadena, CA, USA}

\author[0000-0002-3212-7802]{Patrick O'Brien}
\affil{Lunar and Planetary Laboratory, The University of Arizona, Tucson, AZ 85721, USA}


\begin{abstract}
\kepler's short-period exoplanet population has revealed evolutionary features such as the Radius Valley and the Hot Neptune desert that are likely sculpted by atmospheric loss over time. These findings suggest that the primordial planet population is different from the Gyr-old \kepler population, and motivates exoplanet searches around young stars. Here, we present \pterodactyls , a data reduction pipeline specifically built to address the challenges in discovering exoplanets around young stars and to work with TESS Primary Mission 30-min cadence photometry, since most young stars were not pre-selected TESS 2-min cadence targets. \pterodactyls builds on publicly available and tested tools in order to extract, detrend, search, and vet transiting young planet candidates. We search five clusters with known transiting planets: Tucana-Horologium Association, IC~2602, Upper Centaurus Lupus, Ursa Major and Pisces Eridani. We show that \pterodactyls recovers seven out of the eight confirmed planets and one out of the two planet candidates, most of which were initially detected in 2-min cadence data. For these clusters, we conduct injection-recovery tests to characterize our detection efficiency, and compute an intrinsic planet occurrence rate of 49$\pm$20\% for sub-Neptunes and Neptunes (1.8-6\,\Rearth) within 12.5\,days, which is higher than \kepler's Gyr-old occurrence rates of 6.8$\pm$0.3\%. This potentially implies that these planets have shrunk with time due to atmospheric mass loss. However, a proper assessment of the occurrence of transiting young planets will require a larger sample unbiased to planets already detected. As such, \pterodactyls will be used in future work to search and vet for planet candidates in nearby clusters and moving groups.
\end{abstract}



\section{Introduction} \label{sec:intro}
The \kepler mission was instrumental in the discovery of thousands of exoplanets \citep{Borucki2010, Borucki2017}, most of which were found to be closer to their star than Mercury is to our Sun. A closer look at this population of short-period exoplanets revealed two prominent features: the Hot Neptune Desert and the Radius Valley. The Hot Neptune Desert (3$-$10\,\Rearth{}; $<$4\,days) is a region in the Gyr-old, period-radius plane with a significant dearth of Neptune-sized planets \citep{Beauge2013}. As Neptune-sized planets should be easier to find in short-period orbits, and many sub-Neptunes have been discovered with longer orbital periods from surveys such as \textit{CoRoT} \citep{baglin2006corot} and \textit{Kepler}, the Hot Neptune Desert appears to not be an observational bias. The Radius Valley, a region with a much lower frequency of planets with radii $\sim$1.8\,\Rearth{} than at $\sim$1.3\,\Rearth{} (Super-Earths) or at $\sim$2.4\,\Rearth{} (sub-Neptunes), was discovered by \citet{fulton2017california} using accurate stellar, and hence planetary, radii from the California-\kepler survey \citep{petigura2017california, johnson2017california}. More recently, the Radius Valley was also detected in \ktwo data \citep{hardegree2020scaling}. 

Both the Hot Neptune Desert and the Radius valley are most likely caused by atmospheric mass loss due to XUV photoevaporation (see, e.g. \citealt{owen2013kepler, owen2017evaporation}) and/or core-powered mass loss (see e.g., \citealt{ginzburg2016super, ginzburg2018core, gupta2019sculpting, gupta2020corecool}). Given that these features are most likely to be evolutionary, it is likely that the primordial population of short-period planets was very different. Uncovering that population requires discovering exoplanets around young stars which can be further used to test the theories of atmopheric mass loss.

The \kepler mission, however, was not able to detect any planets younger than 1\,Gyr since there were only four open clusters (0.7-9\,Gyr) in the \kepler field \citep{meibom2011kepler, meibom2013same}. On the other hand, the \ktwo mission \citep{howell2014k2} discovered several young transiting planets while surveying the Taurus, Hyades, Praesepe and Upper Scorpius (e.g., \citealt{mann2016zodiacal_a, mann2016zodiacal_b, mann2017zodiacal, ciardi2018, rizzuto2017zodiacal, rizzuto2018zodiacal, gaidos2017zodiacal, vanderburg2018zodiacal}). Detailed studies of these planets have already begun to shed light on how these planets differ from their Gyr-old counterparts. For example, the planets in the 23\,Myr-old V1298 Tau system, that currently occupy gaps in the \kepler distribution, is thought to be undergoing photoevaporation and is predicted to ``shrink" or lose their atmospheres with time \citep{david2019four}.

With the Transiting Exoplanet Sky Satellite (TESS) mission \citep{ricker2014transiting}, we now have the unique opportunity to detect planets around stars in more young clusters and associations than those observed with \ktwo. During its primary mission, TESS monitored each strip of the sky (also known as sectors) for approximately 27\,days, which means that it can detect planets with an orbital period of $\sim$14 days (assuming two transits per candidate). Per sector, there are about 200,000 bright nearby stars that were pre-selected for the  two year primary mission to be observed in a short 2-minute cadence mode known as the Candidate Target List (CTL; \citealt{stassun2018tess}). Moreover, TESS obtained images of each sector at a 30-minute cadence which are also known as Full-Frame Images (FFIs). Over a billion stars that were observed in this mode are listed as part of the TESS Input Catalog (TIC), which contains a relatively untapped reservoir of young planetary systems. In fact, while a handful of young planets have already been detected using 2-min cadence data within 200\,pc (e.g.,\citealt{newton2019tess,newton2021tess, rizzuto2020tess,mann2020tess}), only two have been detected thus far using the FFIs \citep{nardiello2020psf,bouma2020cluster}.

Automating the detection of young transiting planets poses several challenges, the biggest one primarily being the removal or modeling of instrumental systematics and stellar variability $-$ the latter of which is more important for young stars due to their high variability. Young stars also tend to have stellar rotation rates that are comparable to the orbital periods of short-period planets that can be detected by TESS \citep[e.g.,][]{Rebull2016}, thereby making it hard to find transiting planets in their light curves. A study led by \citet{nardiello2020psf}, \texttt{PATHOS}, used a PSF approach to extract light curves from TESS Primary Mission FFIs and searched for planets in them. However, most of their clusters were far away ($>$500\,pc) and while they were able to find 35 planets, only two of these were within 200\,pc thereby making precise radial velocity of these targets challenging. Additionally, TESS QuickLook Pipeline\footnote{https://archive.stsci.edu/hlsp/qlp} \citep{huang2020photometry_a, huang2020photometry_b, kunimoto2021quick} only considers stars brighter than T mag = 13.5 and is not optimized for highly variable stars such as the ones found in these clusters. To this effect, we have developed a dedicated pipeline to understand the survey completeness and thus properly calculate the intrinsic planet occurrence rates. 

Here, we present \pterodactyls{}\footnote{https://github.com/rachelbf/pterodactyls} (\textbf{P}ython \textbf{T}ool for \textbf{E}xoplanets: \textbf{R}eally \textbf{O}utstanding \textbf{D}etection and \textbf{A}ssessment of \textbf{C}lose-in \textbf{T}ransits around \textbf{Y}oung \textbf{L}ocal \textbf{S}tars), a pipeline that builds on publicly available and tested tools in order to extract, detrend, search, and vet young transiting planet candidates detected using TESS Primary Mission 30-minute cadence photometry. In Section~\ref{sec:sample}, we discuss how we select the clusters that have been included in this analysis and, Section~\ref{sect:PTERODACTYLS} describes the inner workings of \pterodactyls{}. Finally, in Sections~\ref{sec:results} \& \ref{sec:occ_rates}, we summarize our main results and some preliminary occurrence rates calculations followed by a discussion on our next steps (Section~\ref{sec:summary}).


\section{Sample of Young Stellar Clusters} \label{sec:sample}

\begin{table}[!htb]
    \centering
    \begin{tabular}{|c|c|c|c|c|}
    \hline\hline
    Cluster & Distance (pc) & Age (Myr) & Obs/Total\\
    \hline\hline
    THA & $46^{+8}_{-6}$ & 45$\pm$4 & 201/214\\
    IC~2602 & 146$\pm$5 & $46^{+6}_{-5}$ & 502/504\\
    UCL & 130$\pm$20 & 16$\pm$2 & 719/937\\    
    UMa & $\sim$25 & 414$\pm$23 & 16/17\\
    PiEri & 80$-$226 & 120 & 153/254\\    
    \hline\hline
    \end{tabular}
\caption{{Moving groups and clusters that have been homogeneously searched for transiting planets in this paper. Distance, age, and membership are from
\citet{gagne2018banyan} and \citet{babusiaux2018gaia}.} The last column specifies how many stars were observed during the TESS Primary mission vs. the total number of stars in the cluster.}
\label{table:data}
\end{table}

\begin{figure*}[!htb]
   \includegraphics[width=\textwidth]{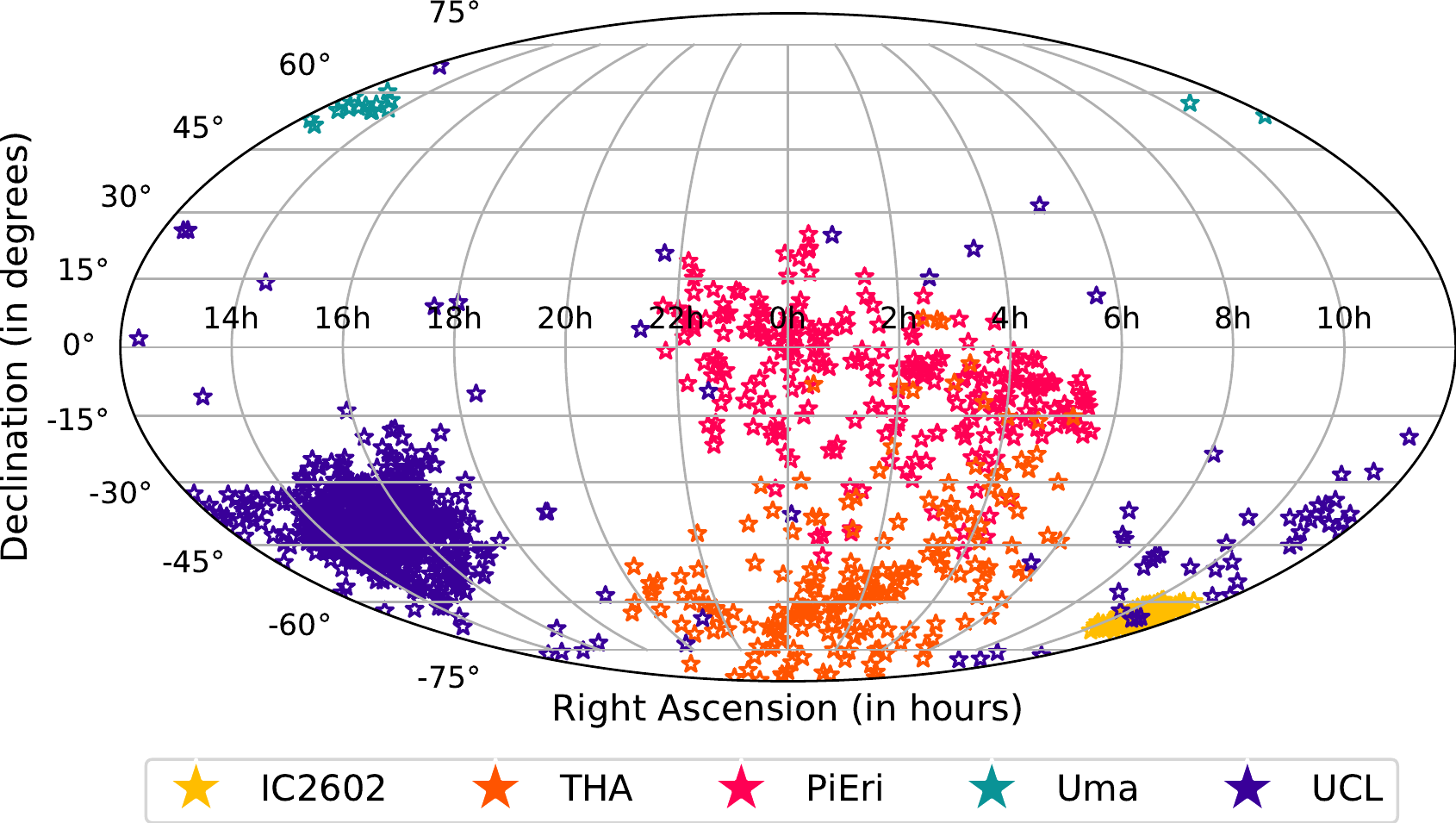}
   \caption{Equatorial locations of stars in selected clusters used in this study} 
   \label{fig:cluster}
\end{figure*}

In order to develop and test \pterodactyls, we limited our search to only those young clusters in which TESS, along with followup observations with the \spitzer space telescope \citep{werner2004spitzer}, had already found transiting planets during its primary mission: Tucana-Horologium Association (THA), IC~2602, Upper Centaurus Lupus (UCL), Ursa Major (UMa), and Pisces-Eridani (PiEri). DS TUC~A\,b \citep{newton2019tess}, HIP~67522\,b and c \citep{rizzuto2020tess}, TOI~1726\,b and c \citep{mann2020tess}, and TOI~451\,b, c and d \citep{newton2021tess} were discovered using TESS 2-min cadence data and also had \spitzer data. All of these are confirmed planets except for HIP~67522\,c. In contrast, the planets in IC~2602 were discovered using 30-min cadence data: TOI~837\,b which is a confirmed planet with mass measurements from radial velocity \citep{bouma2020cluster}, and TIC~460950389\,b \citep{nardiello2020psf} which is a Community TESS Object of Interest (CTOI)\footnote{\url{https://exofop.ipac.caltech.edu/tess/view\_ctoi.php}}. 

In summary, a total of ten planets have been discovered so far in the selected clusters: of them eight were discovered in TESS 2-min cadence data and two in 30-min cadence data. Followup observations have confirmed eight of these while HIP~67522\,c and TIC~460950389\,b are only candidate exoplanets.

For the selected clusters and associations, we relied on the memberships provided in the BANYAN $\Sigma$ \citep{gagne2018banyan} and the \textit{Gaia} DR2 open cluster member lists (Gaia Collaboration, \citealt{babusiaux2018gaia}) which gives us a total of 1940 stars out of which 1591 were observed during the TESS Primary Mission. Based on parameters available in the TIC, these stars span a large range in \gaia G magnitude (1.73-20.45)  and stellar mass (0.11-3.2\,\Msun). Table \ref{table:data} provides a breakdown of the main properties of the moving groups and clusters that were included in this analysis while Figure \ref{fig:cluster} shows the relative positions of the clusters and their respective members that were analyzed in this work. 


\section{Transit Detection Pipeline} \label{sect:PTERODACTYLS}
\begin{figure*}[!htb]
   \includegraphics[width=\textwidth]{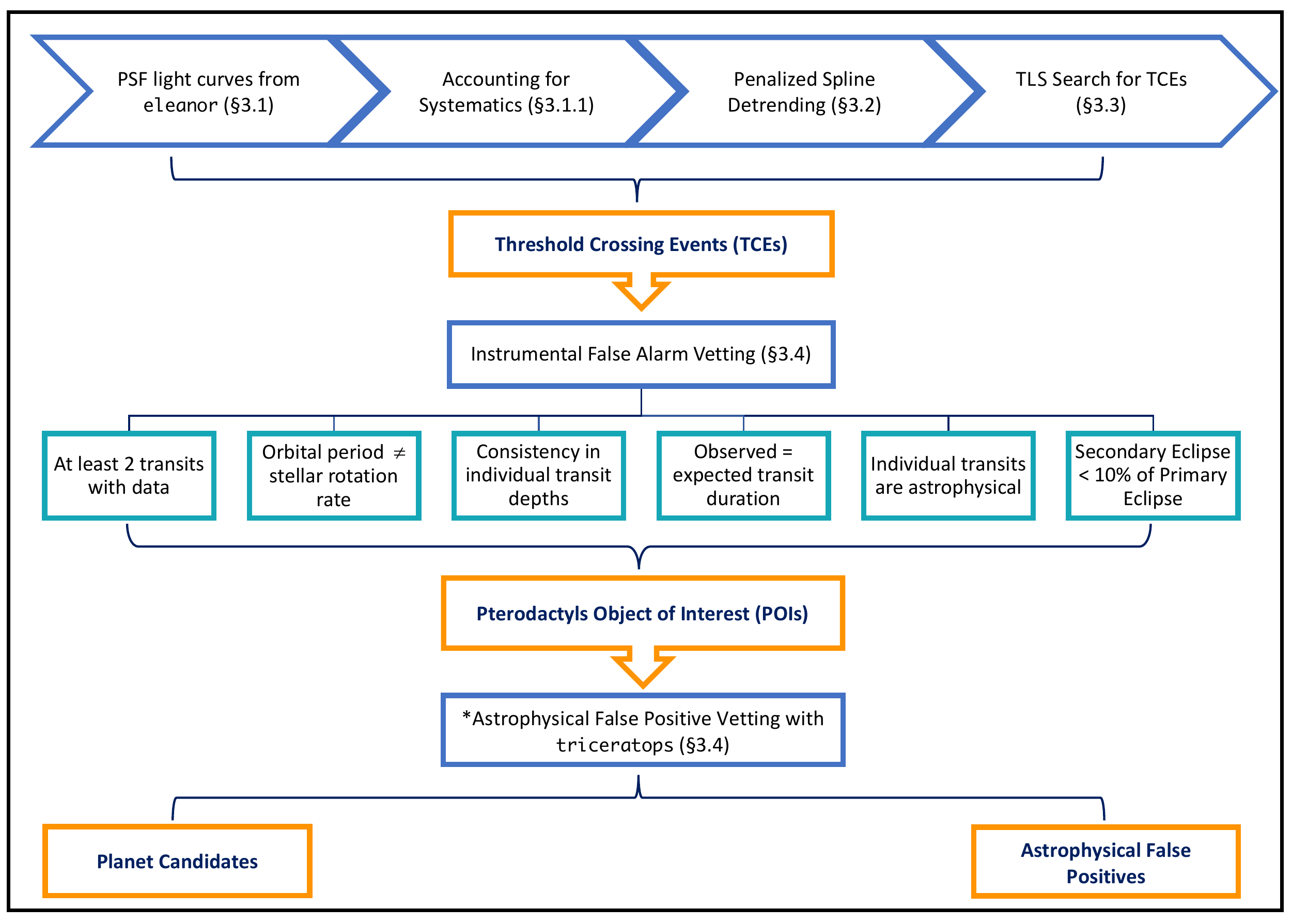}
   \caption{Schematic of \pterodactyls. $^\text{*}$Before passing to \triceratops, we conducted a centroid test whenever the Target Pixel File (TPF) was available on MAST.}
   \label{fig:flowchart}
\end{figure*}

\pterodactyls builds on publicly available and tested tools in order to extract the light curves (Section \ref{subsec:extract}), detrend them (Section \ref{subsec:detrend}), search for transit-like signals (Section \ref{subsec:search}), and vet planet candidates (Section \ref{subsec:vet}) in these clusters. 
We further optimized the efficiency of the detrending, search and vetting components of \pterodactyls{} via injection-recovery tests. For the latter, we injected $\sim$1000 transit signals with planets ranging in size from 0.5-10\,\Rearth{} and with orbital periods between 0.5-10\,days into a sample of our light curves using the Python-based \texttt{batman}\footnote{\url{https://www.cfa.harvard.edu/\~lkreidberg/batman/}} package \citep{kreidberg2015batman}. We then processed the light curves with \pterodactyls and examined the distribution of planets that were recovered. We used this planet distribution as feedback in order to improve \pterodactyls{}. The specifics of the extensive testing using planet injection-recovery tests are summarized in the respective sub-sections while the main steps of the pipeline are summarized in Figure~\ref{fig:flowchart}.


\subsection{Light Curve Extraction}\label{subsec:extract}
Given that less than 10\% of our targets were part of the pre-selected TESS 2-min cadence targets, we use the primary mission's 30-min cadence photometry extracted from TESS FFIs. Here, we employed \eleanor{}\footnote{http://adina.feinste.in/eleanor/}, an open-source tool developed by \citet{feinstein2019eleanor}. \eleanor{} carries out background subtraction, and removal of spacecraft systematics such as jitter and pointing drift. \eleanor provides four different types of light curves: raw light curves, corrected light curves, light curves with systematics removed using Principle Component Analysis (PCA), and Point Spread Function (PSF) modeled light curves.

In this work, we used PSF-modeled light curves since they better enable the recovery of small signals and minimize the effects of scattered light from the Earth and Moon \citep{feinstein2019eleanor}. \eleanor creates these PSF light curves by modeling the shape of the PSF as a multi-variate Gaussian (major axis, minor axis, and a rotation) for each star at each cadence. Then, \eleanor integrates that Gaussian function to calculate the flux of the target star at each cadence. The quality flags assigned to the FFI data by the TESS team are implemented during the extraction of the light curves by \eleanor (see Table 32 in the TESS Science Data Products Description Document\footnote{\url{https://archive.stsci.edu/files/live/sites/mast/files/home/missions-and-data/active-missions/tess/_documents/EXP-TESS-ARC-ICD-TM-0014-Rev-F.pdf}}). By default, we only keep the data with the bit value 0 i.e., the highest quality data. Once we download the light curves for each sector for a particular target, we ‘stitch’ them together and run the search on all available Primary Mission data for the target.


\subsubsection{Accounting for Systematics}
Any instrumental systematics that affect the shape of the PSF, such as the temperature of the instrument, are well-captured by the PSF model and the flux is conserved across it. However, there were several instrumental effects in these light curves that were not captured during the PSF modeling and remain in the light curve such as mid-sector flux drops caused by data downlink, and offsets in the light curves caused by errors in the uploaded Guidestar tables (see TESS' Data Release Notes\footnote{\url{https://archive.stsci.edu/tess/tess_drn.html}} for a more detailed description of these issues). Since these instrumental effects would  generate spikes in the light curve that affect \pterodactyls' ability to detrend and search for planets, we further optimized these light curves by masking the mid-sector flux drops and re-scaling any offsets. In some cases, \eleanor could not properly fit the PSF flux of the target star possibly due to the target being faint, nearby bright stars, etc. We removed 148 such light curves from our sample.

\begin{figure*}[!htb]
  \includegraphics[width=\textwidth]{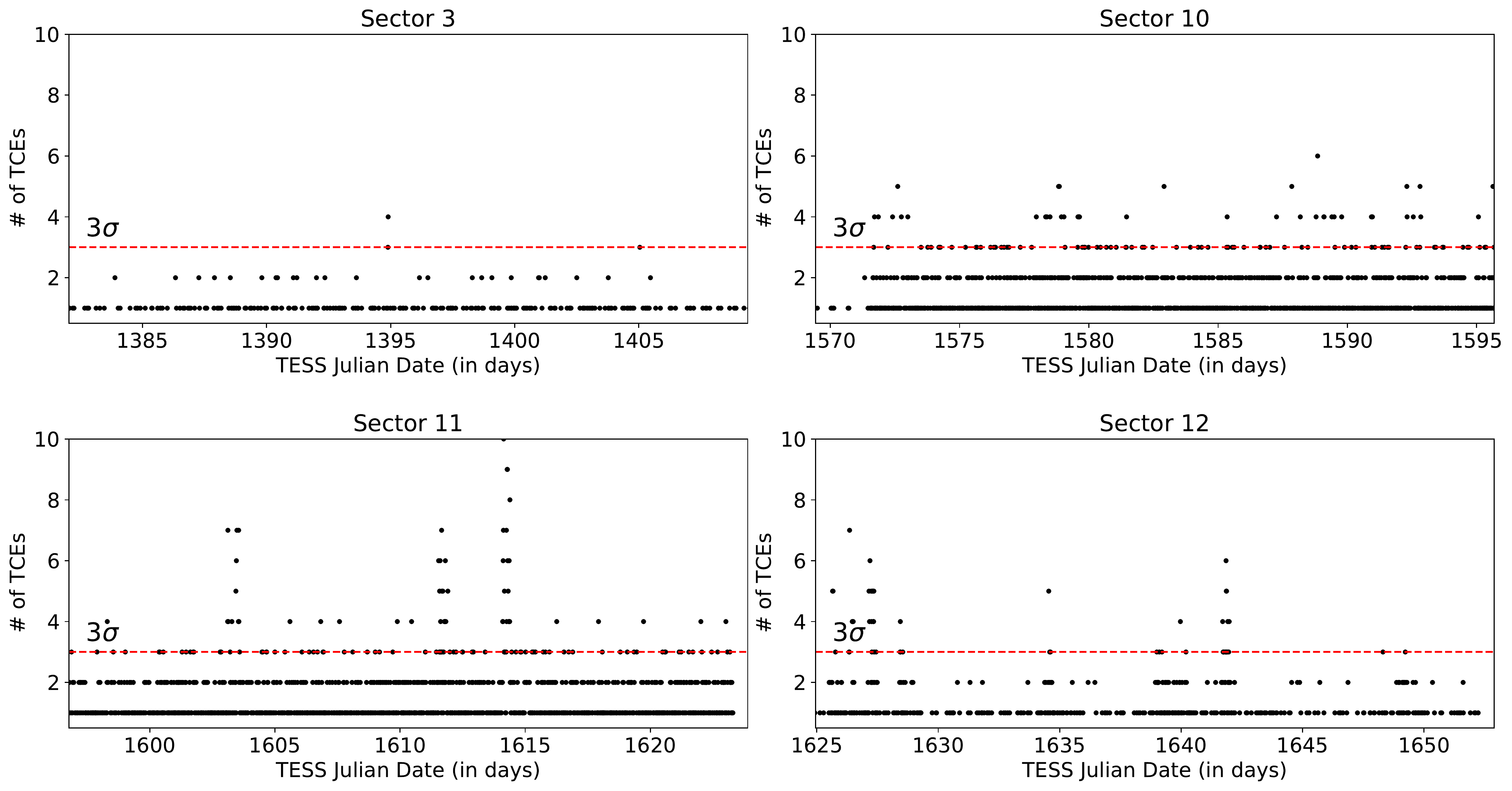}
  \caption{Skye Excess Metric for Sectors 3, 10, 11 and 12 respectively. Red dashed line depicts the 3$\sigma$ threshold.}
  \label{fig:sem}
\end{figure*}

We also masked a number of cadences in sectors 3, 10, 11, and 12 (see Figure~ \ref{fig:sem}) because we noticed that they triggered an anomalous number of transit-like detections in the search, indicating an issue with the light curves in those regions. We used a test similar to the ``Skye" metric used by \citet{thompson2018planetary} that was also adopted for \ktwo data by \citet{zink2020scaling}. The test measures the number of transit-like signals at each cadence and if the number of signals is $>$3$\sigma$, we masked those cadences before re-running the search.

\subsection{Light Curve Detrending}\label{subsec:detrend}
In order to detrend the light curves of these young stars, we explored several options presented in \wotan, an open-source Python package on GitHub\footnote{\url{https://github.com/hippke/wotan/}} developed by \citet{hippke2019wotan}. In their study, \citet{hippke2019wotan} used injection recovery tests and found that, in the presence of significant stellar variation, spline-based methods (where adjacent functional pieces join each other at fixed knots) maximized the fraction of injected planets that were recovered. Following up on this result, we tested the following three different spline techniques (see Appendix~\ref{sec:detrend_test} for an example): i) a robust spline with iterative sigma clipping, ii) a Huber estimator spline \citep{huber1981robust} and, iii) a penalized spline with iterative sigma clipping \citep{eilers1996flexible}. We also tested the commonly used Savitzky-Golay filter \citep{savitzky1964smoothing} and found that it often over-fits for the transit thereby causing an inaccurate transit depth measurement, as pointed out in \citet{hippke2019wotan}. 

We evaluated the detrending based on the low noise or the root mean square (hereafter, RMS) of the light curve, and recovery fraction from the injection-recovery tests. We found that the penalized spline with iterative sigma clipping works best for our sample of young stars as it gives the lowest RMS measured on a 7-hour window and had the highest recovery fraction ($\sim$92\%). We found that, in agreement with \citet{nardiello2020psf}, the brighter stars contribute more flux and have low RMS. We also see a lot of scatter in the faint star region where there is more flux contamination. Furthermore, the ratio of the recovered planetary radii to the injected radii was close to unity, thereby suggesting that we did not over-detrend our light curves.

In carrying out these tests, we also noticed that the edges of light curves are often challenging to detrend due to the lack of data on one side and would often trigger a search. As such, we decided not to consider 0.5\,days worth of data on either edge. We also noticed that the penalized spline uses a cut of 2$\sigma$ i.e., it uses sigma clipping in estimating the trend and does not consider any data that is more than 2$\sigma$ away. While this works well for almost the entire sample, it does not do a good job for light curves with high amplitude ($>$2-3\%) rotational signals where it clips out peaks which leaves periodic residuals that are picked up by the search. We manually identified 20 such light curves and applied a 3$\sigma$ cut to them in order to achieve a better detrending.

\subsubsection{Optimization Based on Stellar Rotation Rates}
A penalized spline is essentially a piecewise polynomial fit in which the routine tries all numbers of splines up to a set maximum, and uses a penalty function to evaluate if a better fit outweighs the added degrees of freedom. The penalized spline provides additional flexibility: unlike other spline methods where the exact number of knots must be chosen manually, the penalized spline identifies the best number of knots by comparing it against the smaller residuals from an improved fit starting from a given maximum number. This feature turns out to be useful for our sample as some young stars are rapid rotators ($<$1\,day) with periodic signals on timescales similar to the orbital period. Using injection-recovery tests, we found that the highest planet detection efficiency could be achieved by limiting the maximum number of splines based on the stellar rotation rate. 

To this effect, we used a Lomb-Scargle periodogram (\citealt{lomb1976least, scargle1982studies}; implemented in \texttt{astropy}) to first calculate the stellar rotation rate and then implement the following step function to determine the maximum number of splines that should be used in order to best detrend any given light curve: 

\begin{itemize}
    \item if \rotrate $\leqslant$ 2 days: max splines = 200
    
    \item if \rotrate $>$ 2 days and $\leqslant$ 10 days: max splines = 200/\rotrate
    
    \item if \rotrate $\geq$ 10 days: max splines = 25
    
\end{itemize}
If the periodogram was not able to find a period, it defaults to the length of the segment (half a sector since there is a break in the middle). Since that is $\sim$10 days, only a small number of splines (25) is used which is ideal for a light curve of low variability. However, this is a \textit{maximum} number of splines which means that the detrending can still pick the right number of splines even if the rotation period is estimated to be shorter than it really is.

At this stage, we visually inspected all of the light curves and found 80 with variablility trends that were difficult to detrend e.g., extremely fast rotators with a rotational rate $<$0.5\,days\footnote{Based on the stellar rotation rate, we ideally would use $\sim$500 splines to detrend the light curve but we realized that by doing so, we would likely be fitting noise.}. Since we were not able to properly detrend these light curves, we also would not have been able to find planets in them. As such, we removed these light curves from our sample and are left with $\sim$1359 light curves that we used to search for planets.


\subsection{Planet Candidate Search}\label{subsec:search}
We searched the detrended light curves for periodic, transit-like signals with \texttt{transitleastsquares} (hereafter, \tls; \citealt{hippke2019optimized}). \tls is optimized for signal detection of small planets by taking into account the effects of limb darkening to better model the ingress and egress shape \citep{manduca1977limb, mandel2002analytic}. \tls searches for these signals by phase-folding the light curve over a range of epochs, transit durations and orbital periods calculated using a function based on the mass and radius of the host star as detailed in \cite{ofir2014optimizing}. The period search grid is set by \tls by dividing the length of the light curve by the minimum number of required transits which is set to two.

For the purposes of our work, we defined a Threshhold Crossing Event (TCE) as having a signal-to-noise ratio (SNR) of 7 and a signal detection efficiency (SDE) of 7\footnote{SDE is essentially the SNR of the signal in Fourier space. For more details, see \citet{kovacs2002box}}. Most transit survey studies choose an SDE value between 6 and 10 (see e.g., \citealt{dressing2015occurrence, livingston2018sixty, siverd2012kelt, pope2016transiting, aigrain2016k2sc, feliz2021nemesis}). While a lower SDE would mean higher completeness, a higher SDE would reduce the number of false positives. Our chosen SNR and SDE cuts reduced the number of false positives, which arose from detrending artifacts caused by the significant ($\sim$20\%) number of rapid rotators with periodic signals comparable to stellar rotation rates. Additionally, an SNR and SDE cut of 7 recovers eight out of the ten the previously detected planets (see Section \ref{sub:known}).



\subsection{Candidate Vetting} \label{subsec:vet}
Transiting surveys (like \kepler, \ktwo and TESS) that target a large number of stars face a major challenge while distinguishing false alarms (transit-like events caused by instrumental artifacts) and astrophysical false positives (transit-like events that are astrophysical in origin but are not caused by planets). The process of sorting out false alarms and false positives from real transit signals is known as vetting. In order to filter out any instrumental false alarms that could be due to detrending artifacts, we used criteria inspired by \texttt{RoboVetter} (\kepler; \citealt{thompson2018planetary}) and \texttt{EDI-Vetter} (\ktwo; \citealt{zink2019edivetter, zink2020scaling}) to create four checks in order to address the issues unique to our sample. Only the TCEs that pass the following tests are considered objects of interest: 

\begin{itemize}[leftmargin=*]
\item \textbf{At least two transits with data:} 
We required two transit signals with in-transit data points in a given light curve which allowed us to find planets with orbital periods of up to 13.5\,days around most of the stars (assuming each star only had single sector data). If we required at least three unique transits like \kepler did, we would be limited to transit signals with orbital periods less than 9\,days.

\item \textbf{Orbital period dissimilar to stellar rotation rate:} 
While analysing the results of the search, we often found residuals at orbital periods similar ($\sim$20\%) to the stellar rotation rate (and some of these at half the rotation rates), despite detrending. To combat this, we flag any such TCEs.

\item \textbf{Consistency in individual transit depths:} 
We found that, quite often, false alarms have transit depths with larger variance than real planets. Similar to \kepler's Robovetter Single Event Statistic (SES), we flagged signals with large variations in transit depth as likely false positives. We weed out these signals based on the ratio of the standard deviations of the individual transit depths to the average transit depth, which we set to be greater than 1.5 for a false positive. This cutoff was determined by preserving known planets while keeping the number of false positives low.

\item \textbf{Observed transit duration should be consistent with expected transit duration:} 
On visual inspection, we found that some TCEs had transit durations that were inconsistent with those of planetary transits. For example, stellar variability can lead to dips with a duration longer than a transit, and single occurrence outliers can appear to have a shorter transit duration. In such cases, it is beneficial to compare the observed transit duration to the expected transit duration that is determined using the planetary parameters calculated by \tls. If the observed transit duration is within 75\% of the expected transit duration, the TCE passes this test. This cut was also set to preserve known planets and reduce the number of false positives.
\end{itemize}

In addition, we also used two checks from \texttt{EDI-Vetter Unplugged} \citep{zink2020scaling} that were initially designed to vet transits in \ktwo data but has now been adapted to vet TESS light curves (for e.g., see \citealt{feliz2021nemesis}).

\begin{itemize}[leftmargin=*]
\item \textbf{Individual Transit Test:} 
Derived from the Marshall test designed by \citet{mullally2016identifying} for \kepler, this test used transit models and those of common non-transiting false positives and fit these to the individual transits. It then used the Bayesian Information Criterion (BIC) to gauge the goodness of these fits, thereby flagging the non-transiting signals that are often assumed to belong to transiting planets.

\item \textbf{Secondary Eclipse Test:} 
This test looked for transit events between the 40-60\% phase range of the folded light curve, allowing for some mild eccentricity variations.  If a meaningful secondary eclipse was identified i.e., the secondary eclipse has a transit depth $>$10\% of the primary transit, this flag was triggered and the transit was labeled a false positive. Secondary eclipses with less than 10\% depth can be associated with reflected light from hot Jupiters.
\end{itemize}

Once we removed the clear false alarms and only have transit-like signals, we were left with 21 \pterodactyls Objects of Interest (hereafter POIs; see Tables \ref{table:known_planets} and \ref{table:binaries}) on which we conducted a centroid test \citep{hedges2021vetting}\footnote{https://pypi.org/project/vetting/} in order to determine whether the transiting signal was indeed coming from the target star (see Appendix~\ref{sec:centroid}). This test compares the distributions of centroids for the data inside and outside the transit. An offset would indicate that the transiting signal was coming from a contaminating source such as a nearby or background star. We chose a probability \textit{p} of 5\% for the Student t-test below which the centroids positions inside and outside transits are not considered to come from the same parent distribution.

All POIs were then passed on to \triceratops\footnote{https://github.com/stevengiacalone/triceratops} \citep{giacalone2020vetting} $-$ a vetting tool designed to distinguish bona fide planets from astrophysical false positives in TESS data. \triceratops calculated the marginal likelihood of each transit-producing scenario using observational constraints such as the transit depth and transit duration from \tls for each potential host, positions and magnitude of nearby stars from \gaia \texttt{dr2} \citep{gaiadr2}, priors on planet occurrence and stellar multiplicity rates, and models of planetary transits and eclipsing binaries. These scenarios include transiting planets and eclipsing binaries originating from the target star, known nearby stars, and unknown unresolved stars located near the target on the sky. Lastly, \triceratops{} calculates the Bayes factor between all scenarios involving a transiting planet around the target star and all astrophysical false positive scenarios. Using this Bayes factor, \triceratops can place a statistically robust constraint on the reliability of any given planet candidate detected by TESS. Here, we made a transiting planet probability cut of 50\% in order to preserve known planets.

\section{Fitting for Planet Properties}
Once the POIs have passed all of the vetting tests described in Section \ref{subsec:vet}, we fit the phase-folded light curve\footnote{Since the PSF light curves from \eleanor do not currently have uncertainties, we use the RMS of the detrended light curve as errors while fitting for planet parameters} in order to derive a more accurate planet radius. Since \texttt{TLS} only provided an initial fit to the phase-folded light curve and did not provide uncertainties on the planet radius, we used \tls' outputs as priors for a more rigorous optimization in order to derive uncertainties using \texttt{EXOTIC} \citep{zellem2020utilizing}. Here, the transit parameters are optimized using the multi-modal nested sampling algorithm called \texttt{UltraNest} \citep{buchner2021ultranest} which is a Bayesian inference tool that uses the Monte Carlo strategy of nested sampling to calculate the Bayesian evidence allowing simultaneous parameter estimation and model selection. A nested sampling algorithm is efficient at probing parameter spaces which could potentially contain multiple modes and pronounced degeneracies in high dimensions; a regime in which the convergence of traditional Markov Chain Monte Carlo (MCMC) techniques becomes incredibly slow (\citealt{Skilling2004}; \citealt{Feroz2008}). This approach has been validated on \textit{Kepler} light curves, TESS data and ground-based measurements \citep{Pearson2018, Pearson2019a, Pearson2019b}.


\section{Results} \label{sec:results}
\begin{figure*}[!htb]
\minipage{0.5\textwidth}
  \includegraphics[width=\linewidth]{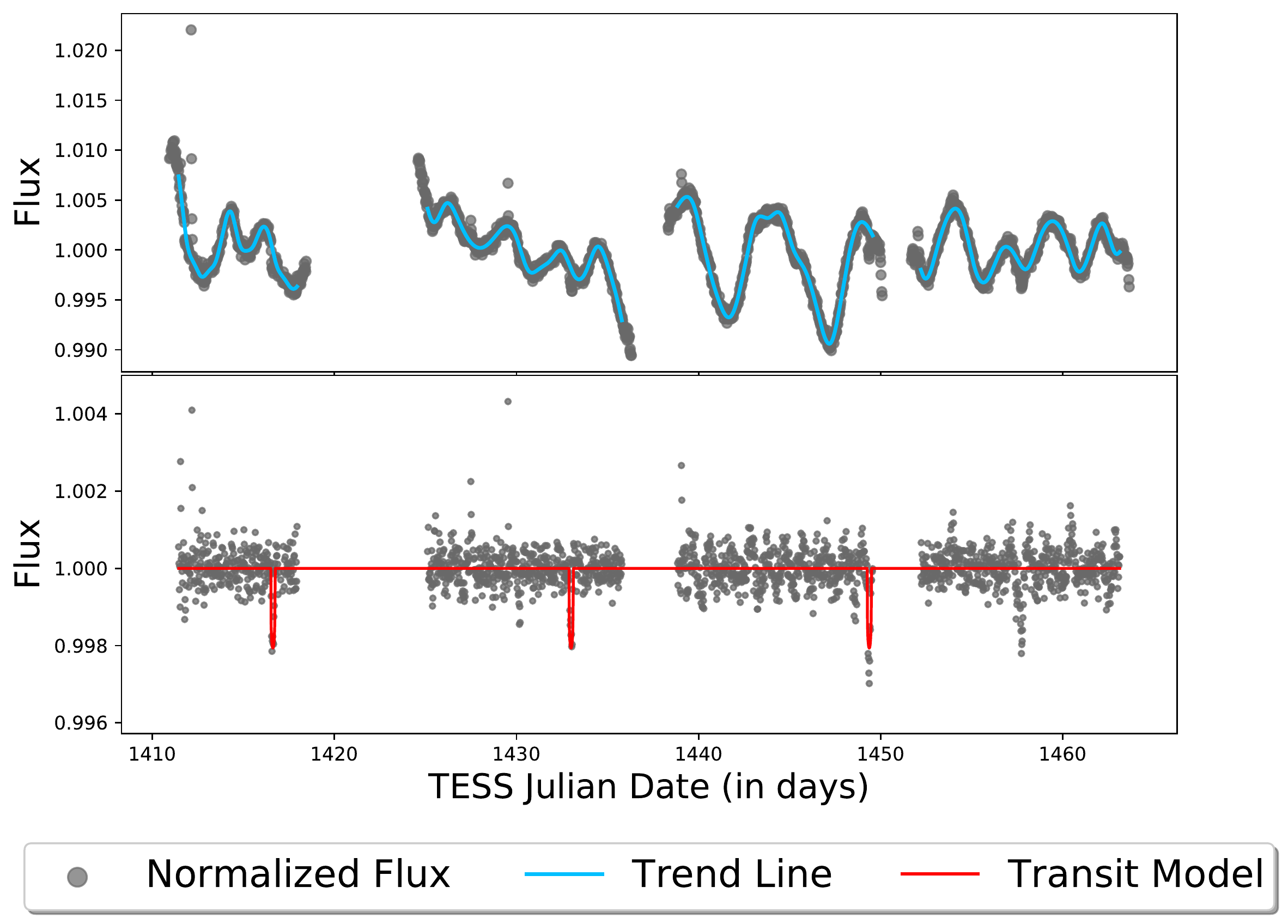}
\endminipage\hfill
\minipage{0.5\textwidth}
  \includegraphics[width=\linewidth]{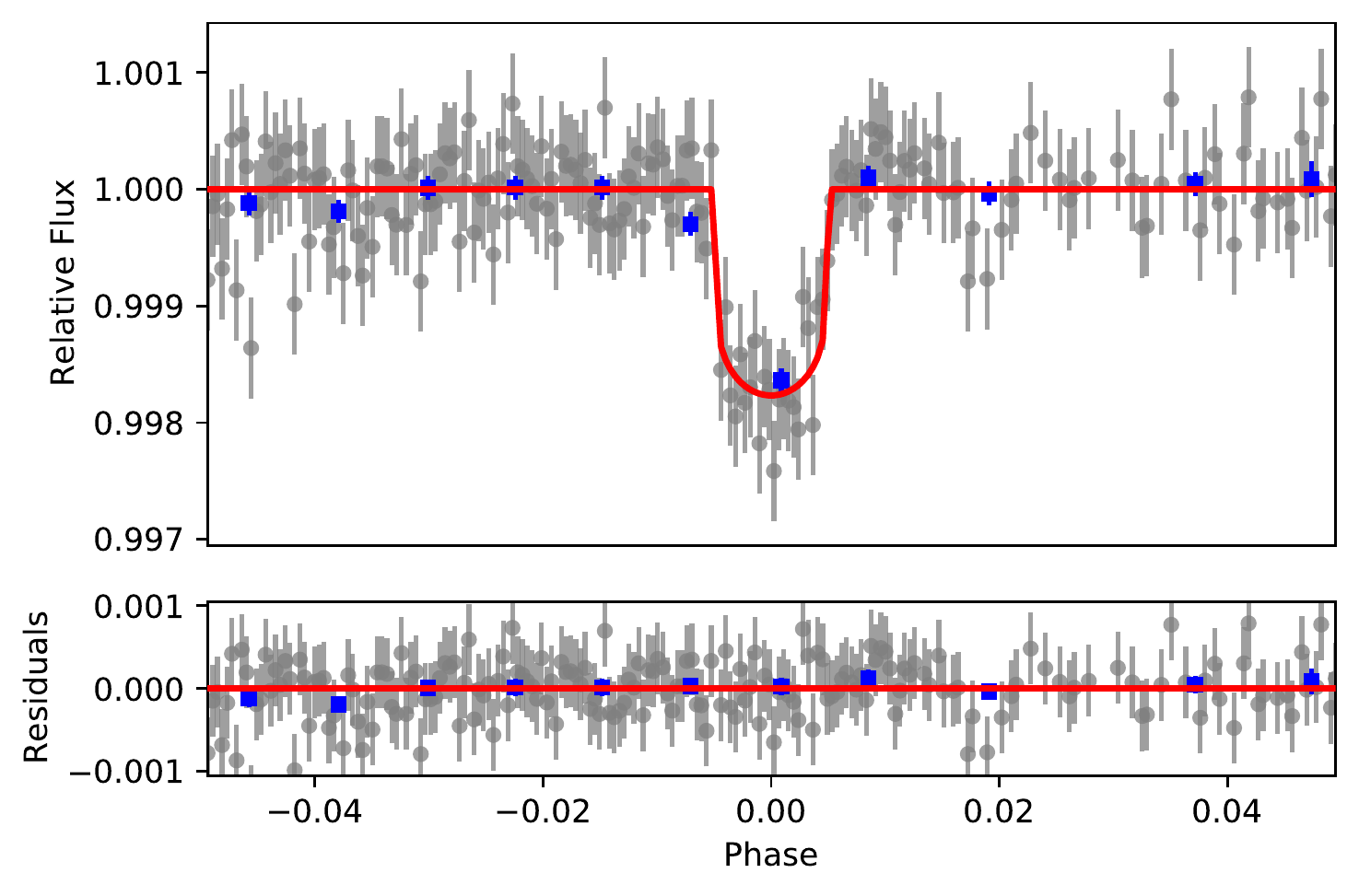}
\endminipage
\caption{\textbf{Left:} Recovery of TOI~451\,d. Upper panel: PSF flux (in dark gray) and penalized spline trend (in blue). Lower panel: Detrended light curve (PSF flux divided by the trend). The multi-transit signature (in red) can be clearly seen. \textbf{Right:} Upper panel shows Phase-folded light curve of TOI~451\,d (gray data points) along with the best fit model using \exotic (in red). Lower panel shows the residuals in parts per million.}
\label{fig:pieri_d}
\end{figure*}

We processed $\sim$1359 young stellar light curves extracted from TESS FFIs i.e., 30-min cadence data through \pterodactyls (see Appendix~\ref{sec:tics} for the TIC IDs processed in this work). As an illustration, Figure~\ref{fig:pieri_d} shows the different steps that were used in the recovery of the planet TOI~451\,d. We detected 21 POIs that were then passed to \triceratops where their phase-folded light curves were vetted to separate transiting planet signals from astrophysical false positives.

\subsection{Recovery of Known Exoplanets} \label{sub:known}

\begin{table*}[!htb]
    \centering
    \begin{tabular}{|c|c|c|c|c|c|c|c|} 
    \hline\hline
Planet & Ref. & Cluster & Radius (pub; \Rearth) & Period (pub; days) & Radius (rec; \Rearth) & Period (rec; days) & TP prob \%\\ 
 \hline
DS Tuc A\,b$^{*}$ & a & THA & 5.70$\pm$0.17 & 8.138268$\pm$0.000011 & $5.349^{+0.464}_{-0.458}$ & 8.215$\pm$0.066 & 54.93\\

TIC~460950389\,b & b & IC~2602 & 3.8$\pm$0.2 & 2.8622$\pm$0.0004 & $3.554^{+0.320}_{-0.304}$ & 2.861$\pm$0.005 & 99.69\\

TOI~837\,b & b,c & IC~2602 & $6.9^{+0.6}_{-0.4}$; $8.90^{+0.74}_{-0.71}$ & 8.3252$\pm$0.0006 & $7.224^{+0.410}_{-0.387}$ & 8.325$\pm$0.016 & 97.45\\

HIP~67522\,b$^{*}$ & d & UCL & $9.72^{+0.48}_{-0.47}$ & 6.95993$\pm$0.00034 & $13.400^{+1.354}_{-1.328}$ & 6.893$\pm$0.019 & 86.06\\

HIP~67522\,c$^{*}$ & d & UCL & $8.01^{+0.75}_{-0.71}$ & $54^{+70}_{-24}$ & xx & xx & xx \\

TOI~1726\,b$^{*}$ & e & UMa & 2.15$\pm$0.10 & $7.10793^{+0.00046}_{-0.00034}$ & $2.134^{+0.214}_{-0.213}$ & 7.116$\pm$0.043 & 75.32\\

TOI~1726\,c$^{*}$ & e & UMa & 2.67$\pm$0.12 & 20.5455$\pm$0.0011 & $2.225^{+0.264}_{-0.267}$ & 20.5234$\pm$0.1022& 75.72\\

TOI~451\,b$^{*}$ & f & PiEri & $1.94^{+0.15}_{-0.34}$ & $1.858701^{+0.000027}_{-0.000033}$ & xx & xx & xx\\

TOI~451\,c$^{*}$ & f & PiEri & 3.07$\pm$0.14  & $9.192523^{+0.000064}_{-0.000084}$ & $2.604^{+0.308}_{-0.294}$ & 9.195$\pm$0.036 & 71.11\\

TOI~451\,d$^{*}$ & f & PiEri & 4.03$\pm$0.15 & $16.364981^{+0.000047}_{-0.000049}$ & $3.678^{+0.306}_{-0.294}$ & 16.368$\pm$0.070 & 70.63\\
\hline\hline
\end{tabular}
    \caption{Published and Recovered Properties of planets detected in young clusters. References: a) \citet{newton2019tess};
    b) \citet{nardiello2020psf}; c) \citet{bouma2020cluster};
    d) \citet{rizzuto2020tess}; e) \citet{mann2020tess}; f) \citet{newton2021tess}
    $^{*}$Planet originally detected in 2-min cadence data.}
    \label{table:known_planets}
\end{table*}

In the five clusters that were searched in this work, there are eight confirmed planets and two planet candidates. More specifically, three were single planet systems (DS Tuc~A\,b, TIC~460950389\,b, and TOI~837\,b; the latter two discovered in 30min-cadence data) and seven were part of three multi-planet systems (HIP~67522\,b and c, TOI~1726\,b and c, and TOI\,451\,b, c and d), all discovered in 2-min cadence data (see Section~\ref{sec:sample}). As \pterodactyls, which is automated, is built to search for just one planet per star, the discovery of additional planets in a system is carried out after finding a first planet and masking it out.

Among the 21 POIs detected by \pterodactyls in the first automated run, we find six of the confirmed exoplanets that were previously detected by various teams using TESS (summarized in Table \ref{table:known_planets}; for light curves, see Appendix~\ref{sec:recovery}). \pterodactyls was able to recover the values for the orbital periods and radii for three of these planets that were consistent (within uncertainties) with the literature. This includes two single planets (TIC~460950389\,b, and TOI~837\,b) that were initially discovered in 30-min cadence data, and one multi-planet system (TOI~451\,b, c and d) that was initially discovered in 2-min cadence data. We first recovered TOI~451\,d (4.1\,\Rearth) at an orbital period of 16.36\,days. After masking this transit, we were also able to recover TOI~451\,c (3.1\,\Rearth) at 9.19\,days (see Figures~\ref{fig:pieri_d} and \ref{fig:pieri_c}). However, \pterodactyls was not able to recover TOI~451\,b (1.9\,\Rearth) since it has a low SNR ($\sim$5). We find that all of the confirmed planet and planet candidates that \pterodactyls recovers pass the centroid test.

\begin{figure*}[!htb]
\minipage{0.5\textwidth}
  \includegraphics[width=\linewidth]{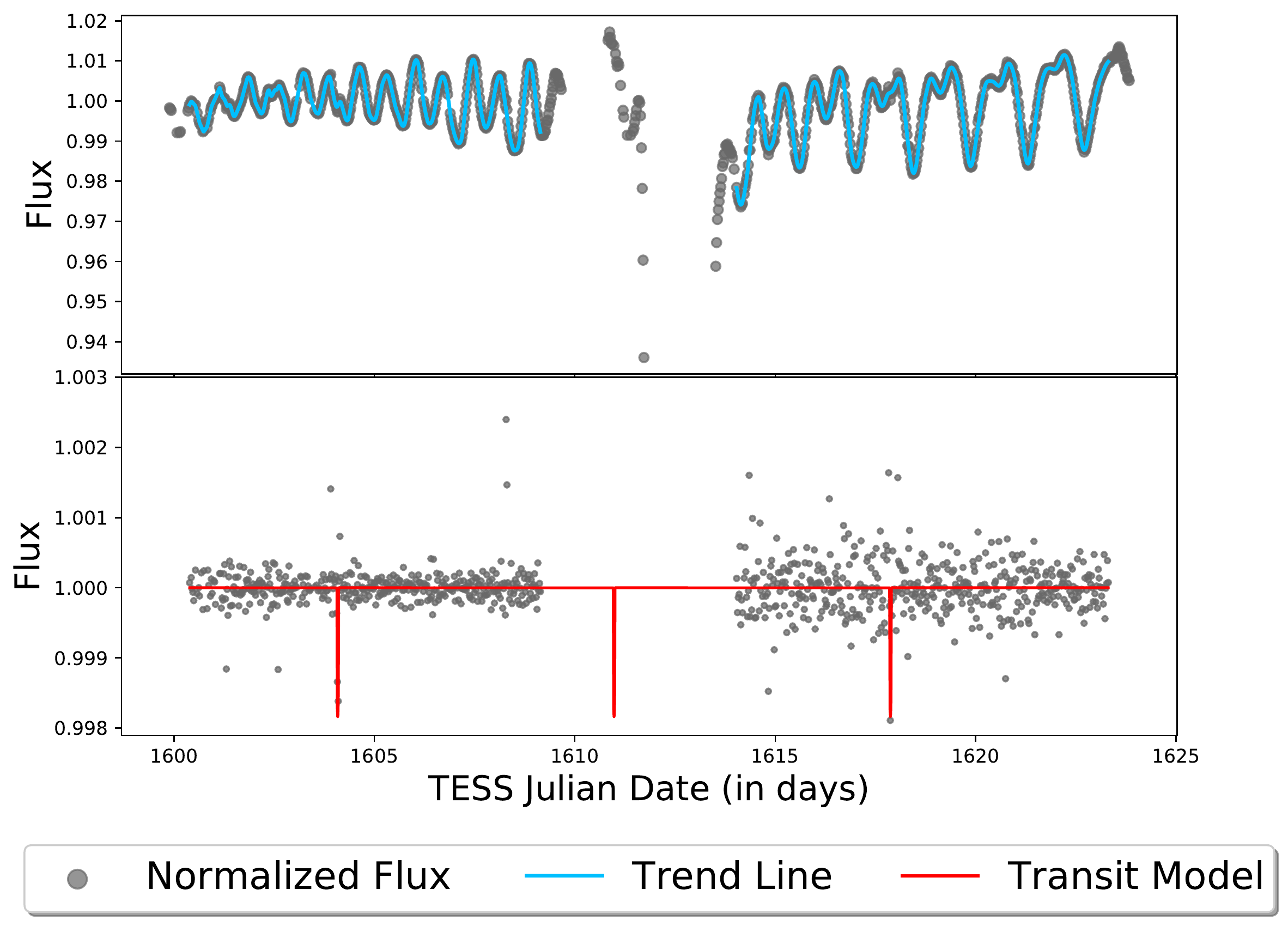}
\endminipage\hfill
\minipage{0.5\textwidth}
  \includegraphics[width=\linewidth]{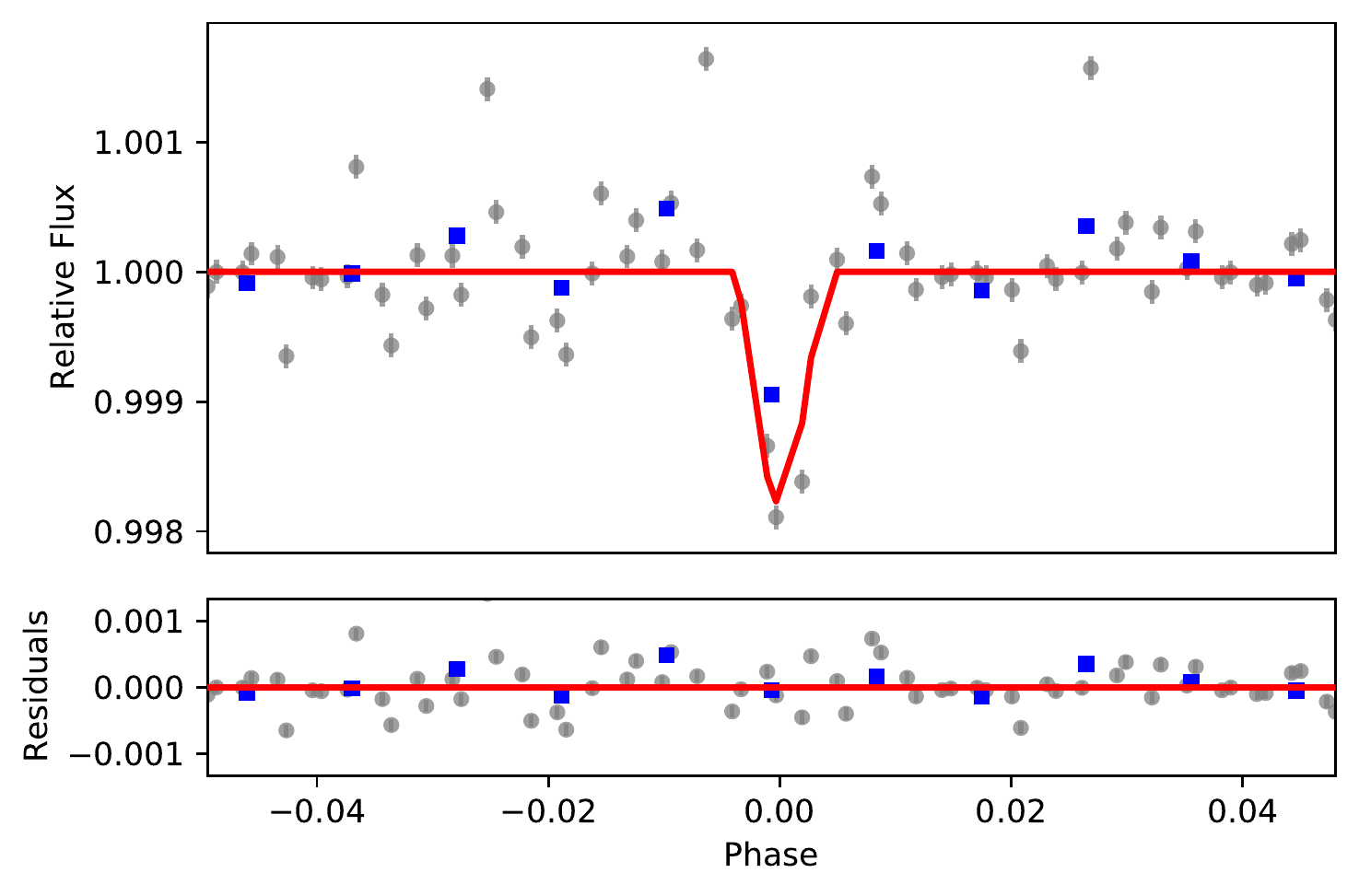}
\endminipage
\caption{\textbf{Left:} Recovery of HIP 67522\,b. Upper panel: PSF flux (in dark gray) and penalized spline trend (in blue). Lower panel: Detrended light curve (PSF flux divided by the trend). The multi-transit signature (in red) can be clearly seen. \textbf{Right:} Upper panel shows Phase-folded light curve of HIP 67522\,b (gray data points) along with the best fit model using \exotic (in red). Lower panel shows the residuals in parts per million.}
\label{fig:ucl_hip67522}
\end{figure*}

\begin{figure*}[!htb]
\minipage{0.5\textwidth}
  \includegraphics[width=\linewidth]{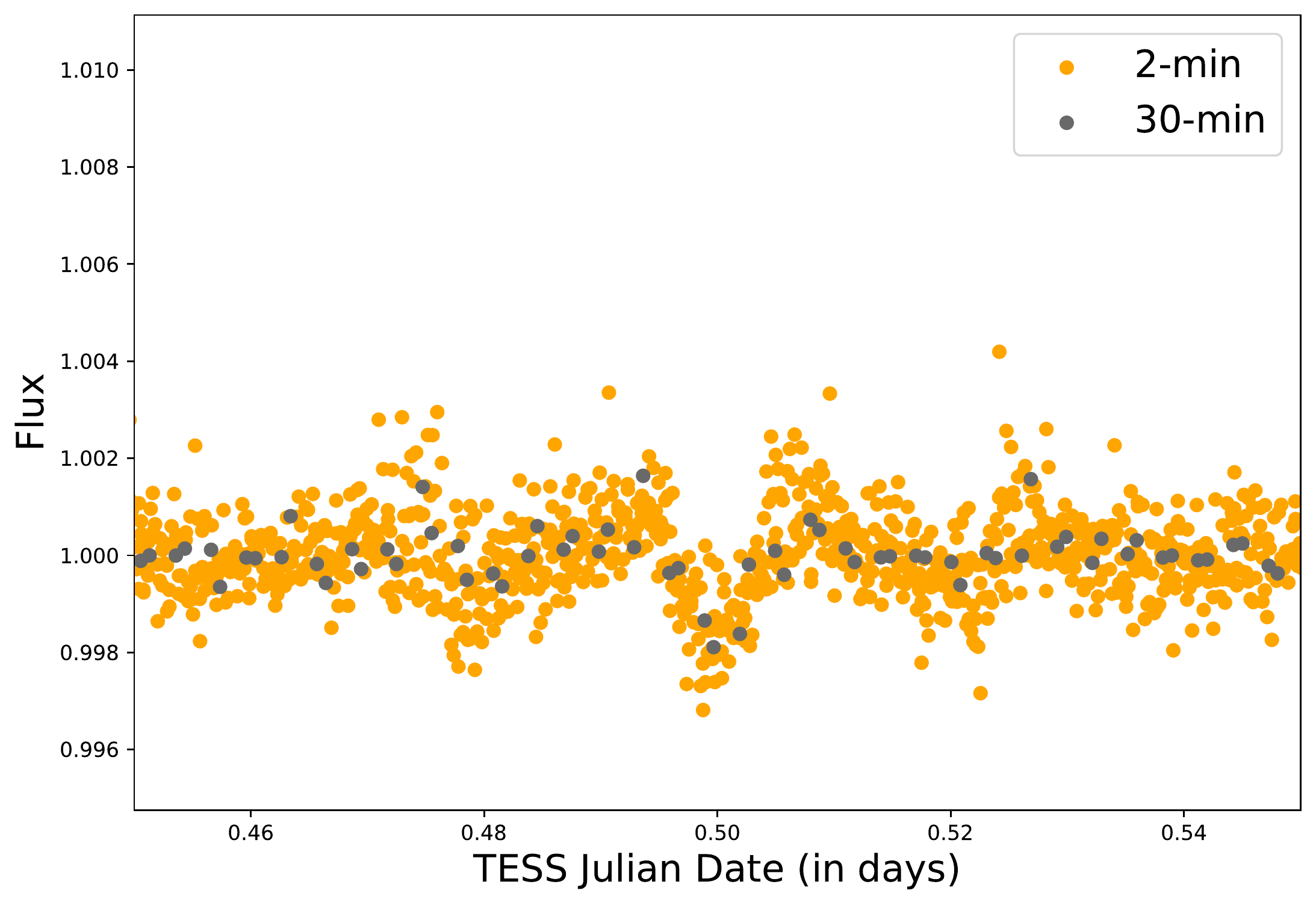}
\endminipage\hfill
\minipage{0.5\textwidth}
  \includegraphics[width=\linewidth]{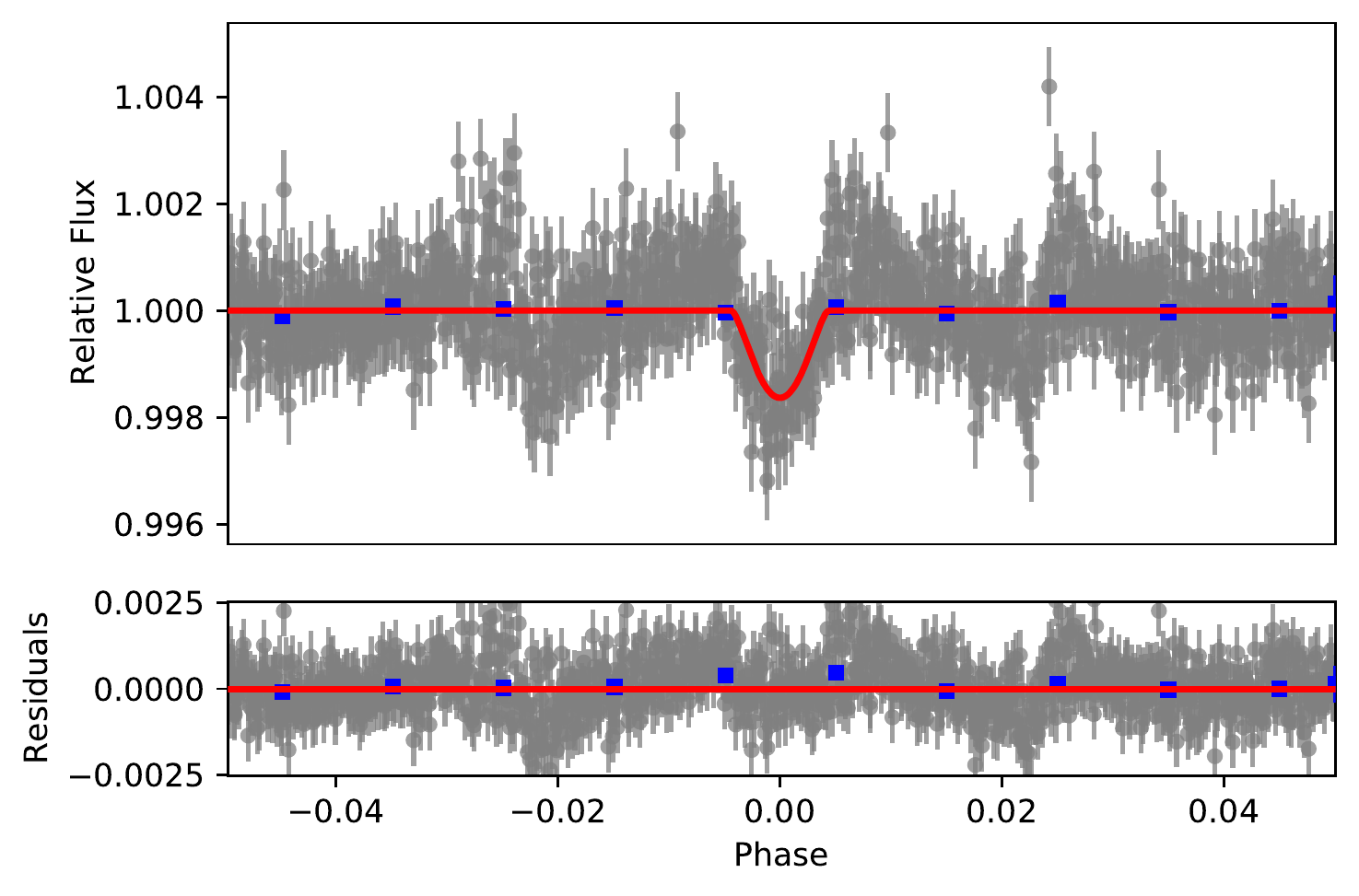}
\endminipage
\caption{\textbf{Left:} Comparison of phase-folded light curve of HIP 67522\,b using 2-min cadence data (orange) and 30-min cadence data (black). \textbf{Right:} Upper panel shows Phase-folded light curve of HIP 67522\,b using 2-min cadence (dark gray points) along with the best fit model using \exotic (in red). Lower panel shows the residuals in parts per million.}
\label{fig:hip67522_2mins}
\end{figure*}

Two out of the other three recovered transiting systems (out of a total of six as stated above) suffered from sparse cadence (few in-transit data points): DS Tuc~A\,b and HIP~67522\,b. For these, \pterodactyls was able to accurately recover the orbital periods but not the same radii as in the literature. In the case of DS Tuc~A\,b, at first we considerably underestimated the planet's radius value. On further inspection, we noticed that this was caused by very few in-transit data points that was due to the removal of ``bad" data that was flagged using the TESS team's quality flags. In this specific case, we found that inclusion of certain quality flags lead to a better estimation of the planetary radius i.e 5.35\,\Rearth (see Figure~\ref{fig:dstuc}) which is within errors of the value of 5.70\,\Rearth that was measured using TESS' 2-min cadence and \spitzer data (see Table~\ref{table:known_planets}). In the case of HIP~67522\,b, the overestimation of the planet radius was severe: we recovered a radius of 13.4\,\Rearth while the planet's observed radius is 9.72\,\Rearth. This could be due to several factors. First, as can be seen in Figure~\ref{fig:ucl_hip67522}, the middle transit was not recovered (but predicted by \tls) due to the default edge cutoff for all stars being 0.5\,days. However, the severe overestimation of the radius could also be due to the planet being discovered in the more abundant 2-min cadence data. We tested this by processing the publicly available 2-min cadence light curve through \pterodactyls (see Figure~\ref{fig:hip67522_2mins}) and found that with the increased number of in-transit data points (along with higher RMS compared to 30-min cadence data), we were able to recover a planet radius of 9.8\,\Rearth which is close to the literature value of 9.72\,\Rearth. We were not able to recover HIP~67522\,c due to it being a single transit event and hence, beyond the scope of \pterodactyls since it requires a minimum of two transits.

Finally, for the TOI~1726 system in UMa, we recovered a transit signal from a 2.48\Rearth\ planet with an orbital period of 11.32\,days. However, the TOI-1726 system has two small planets (b and c) of radii 2.15\,\Rearth and 2.67\,\Rearth, with orbital periods of 7.11 and 20.55\,days respectively \citep{mann2020tess}. Given that the planet sizes (and hence, transit depths) are similar, the signals were initially detected by \pterodactyls as coming from one planet. Upon further inspection, we realized that \tls set the maximum orbital period in the grid based on the minimum number of transits. So, for single sector data ($\sim$27\,days) with the minimum number of transits set to 2, the maximum orbital period would be $\sim$13.5\,days which is shorter than the orbital period of TOI~1726\,c. With this in mind, for this case alone, we set the minimum number of transits to one and were able to recover TOI~1726\,c (see Figure~\ref{fig:uma_c}). We then masked this transit and re-processed it through \pterodactyls and recovered TOI~1726\,b as well (see Figure~\ref{fig:uma_b}).

\begin{figure*}[!htb]
\minipage{0.5\textwidth}
  \includegraphics[width=\linewidth]{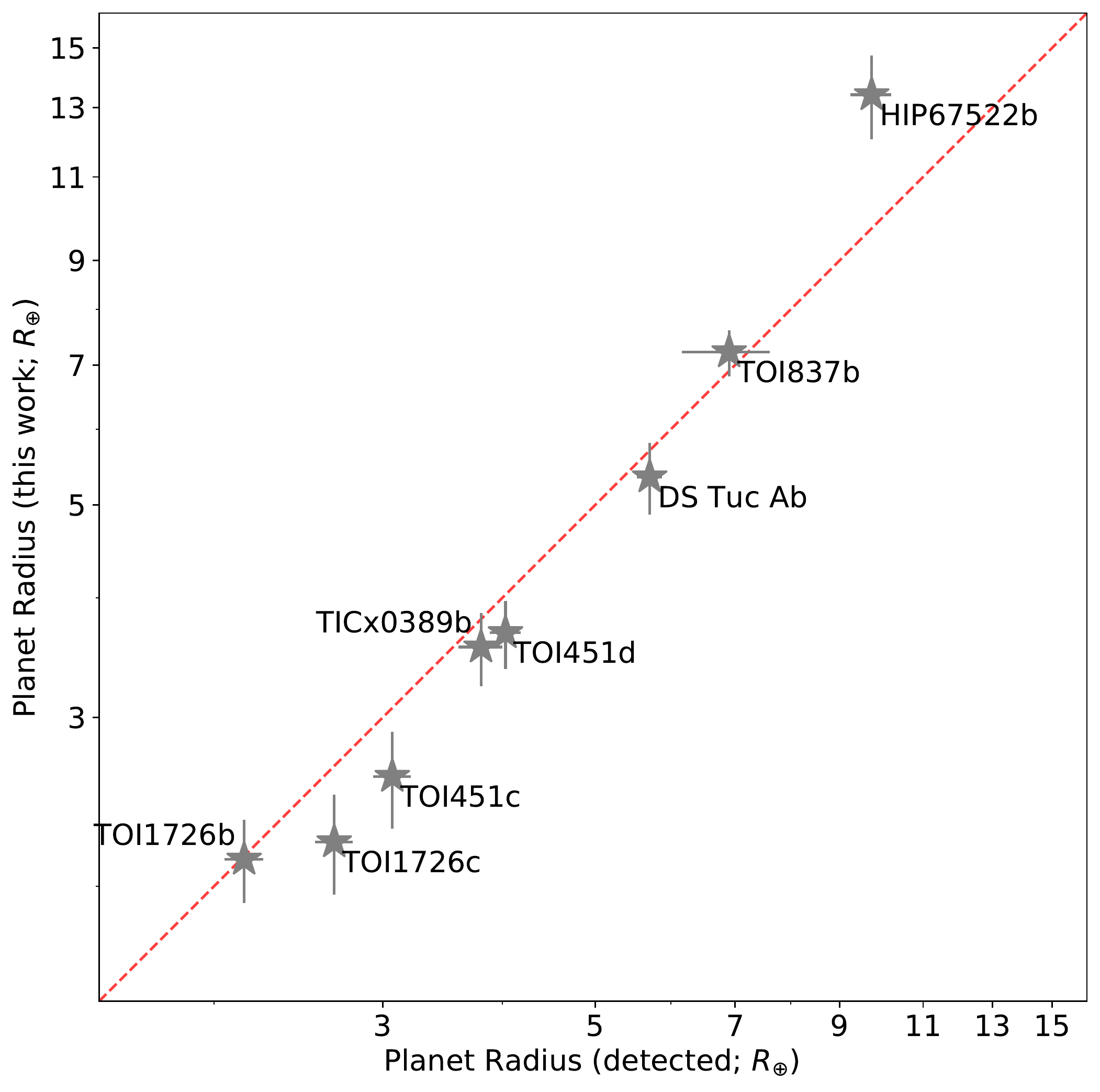}
\endminipage\hfill
\minipage{0.5\textwidth}
  \includegraphics[width=\linewidth]{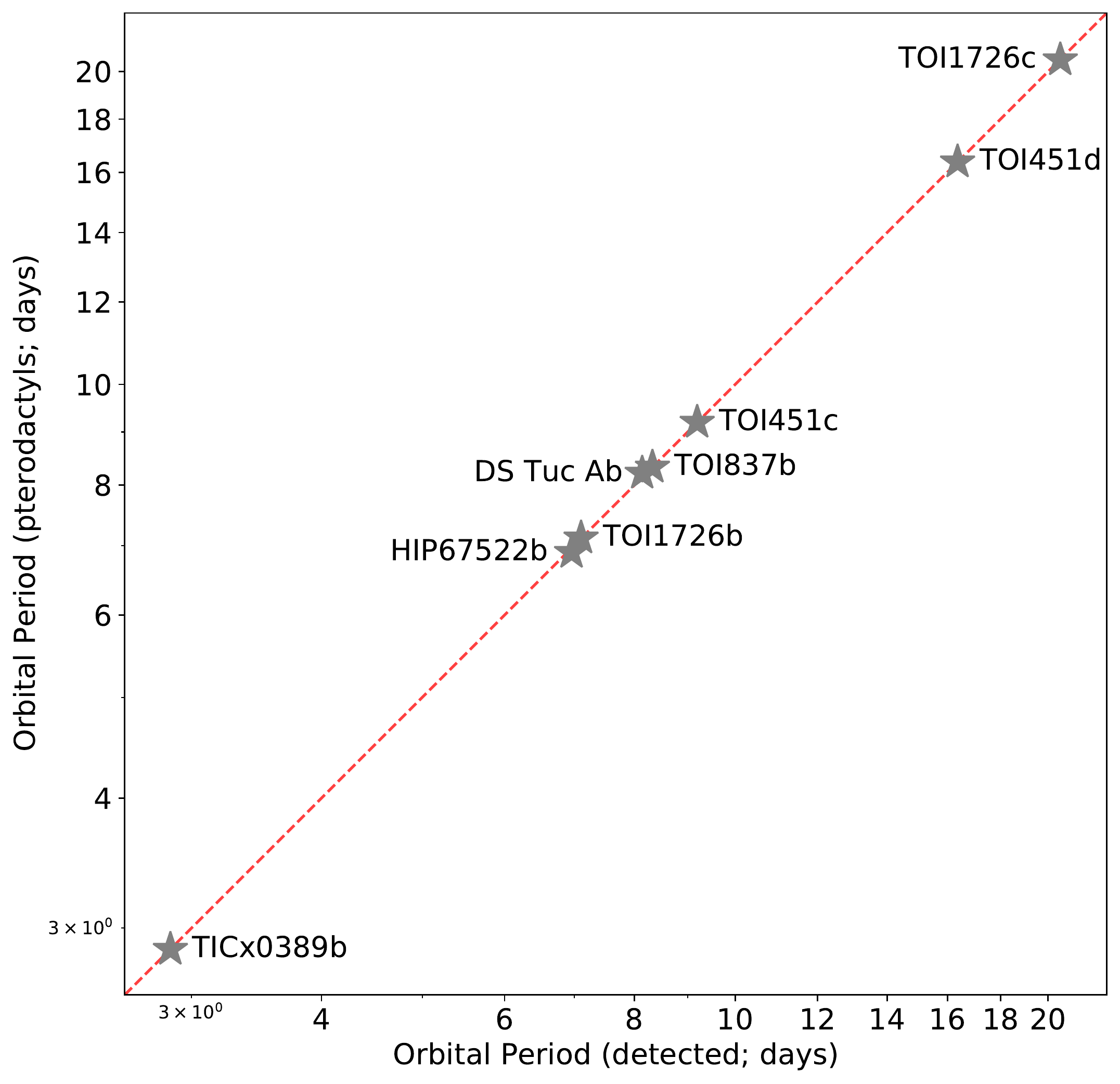}
\endminipage
\caption{\textbf{Left:} Comparison of planetary radii recovered using \pterodactyls (fitted with \exotic) vs. known planet radii for the eight recovered known planets. As can be seen, despite being recovered from 30-min cadence data, the radii are often consistent with those published. In two cases, TOI~1726\,c and TOI~451\,c, we slightly underestimate the planet's radius. We also overestimate the radius of the planet for HIP~6752\,b. \textbf{Right:} Comparison of orbital periods recovered using \pterodactyls vs. known orbital periods for the eight recovered planets to depict consistency between the results of our work and those from the discovery papers. Note: Orbital period errors are extremely small and hence cannot be seen in the plot.}
\label{fig:stellar_comp}
\end{figure*}

After implementing the aforementioned fixes, we find that all of the recovered orbital periods are the same as those reported in the discovery papers while for the radii, four are the same within 1$\sigma$ and three within 2$\sigma$ (Figure~\ref{fig:stellar_comp}). The recovery of these planets demonstrates the usefulness of \pterodactyls in discovering young planets in 30-min cadence data.

\subsection{Astrophysical False Positives Vetting} \label{sub:fps}
Apart from the six confirmed planets, \pterodactyls also detected 15 other POIs (see Table \ref{table:binaries}), including two eclipsing binary signals around two stars in UCL i.e., TIC~129116176 and TIC~148158540. We weren't able to conduct centroid tests on all of these targets since the Target Pixel File on the MAST server (which is used for the centroid test) were only available for three of the candidates (all of the ones in UCL) in Table~\ref{table:binaries} but not for the 12 other candidates (IC2602). As such, we conducted the centroid test whenever the data was available. For the targets, where we could not conduct the centroid test, we visually inspected the eleanor pixel by pixel light curves. While determining the probability of a transiting signal being due an astrophysical false positive or a transiting planet, \triceratops (by default) takes into account planet occurrence rates priors that are based on the Gyr-old exoplanet population (e.g., \citealt{howard2012planet, dong2013fast,petigura2013prevalence,dressing2015occurrence,mulders2015stellar,Mulders2018}). However, if the planet population around young stars is different from that around Gyr-old stars, for example because young planets are larger in size, the probability that \triceratops would calculate is likely to be biased. In order to test the effect of priors on the probabilities of the different scenarios investigated by \triceratops, we also calculated the astrophysical false positive probability using non-informative (flat) priors that were recently implemented in \triceratops.

Out of our remaining 15 POIs, we found that 11 were not affected by the choice of priors and had the same probability of being an astrophysical false positive in both cases. 
In two out of the remaining four, we detected a transiting signal in each each of them (TIC~466579089 and TIC~466035035; see Table \ref{table:binaries}) that had passed all our vetting tests. Using Gyr-old priors, \triceratops calculated the most likely scenario to be a nearby eclipsing binary (NEB). However, when using flat priors, the most likely probability was calculated to be a transiting planet. Given that IC~2602 is an extremely crowded cluster ($\sim$5 stars per TESS pixel), we used \eleanor's pixel level light curves to investigate whether these transits were indeed around our target stars. In doing this, we found that the signal was coming from giant stars (TIC~466579123 and TIC~466035000, respectively) a couple of pixels away and actually contributed more flux to the pixel than our target star. With this information in hand, we ran \pterodactyls on those stars and were able to detect the same signal in the light curve but, this time, \triceratops found that the most likely scenario was that the signal was due to an eclipsing binary around TIC~466579123 and TIC~466035000.

When the false positive probability was calculated using flat priors, the remaining two POIs that were initially calculated to be nearby eclipsing binaries with Gyr-old priors, showed a $>$25\% chance that the signal was either a nearby eclipsing binary or a nearby transiting planet. These POIs have large transits depths that could either be attributed to small stars, brown dwarfs or giant planets since the two populations have comparable radius distributions \citep{chen2017prob}. This, along with the fact that identifying shallow secondary transits can be arduous, makes it harder for statistical validation tools such as \triceratops to properly distinguish between these scenarios \citep{shporer2017three}. In cases such as these, masses obtained via radial velocity follow-up are necessary to differentiate between these scenarios. This illustrates the need to properly take into account flux contamination in young clusters while vetting planet candidates.

\begin{table*}[!htb]
    \centering
    \begin{tabular}{|c|c|c|c|c|c|} 
    \hline\hline
TIC~ID & Cluster & Period (d) & Most likely (Gyr-old priors) & Most likely (Flat priors) & Potential Host Star(s)\\ 
 \hline
390758843 & IC~2602 & 0.890 & 25\% NEB & 25\% NEB & TIC~390758947\\ 

376731139 & IC~2602 & 12.769 & 50\% NEBx2P & 49\% NEBx2P & TIC~376731097\\

911870867 & IC~2602 & 20.018 & 42\% NEB & 40\% NTP; 27\% NEB & TIC~390843430\\

911870847 & IC~2602 & 20.018 & 50\% NEB & 37\%NEB; 30\% NTP & TIC~390843430\\ 

315312718 & IC~2602 & 2.203 & 28\% NEB & 30\% NEB & TIC~315312721; TIC~315312651\\

466035035 & IC~2602 & 1.522 & 99\% EB & 92\% TP$^{x}$ & TIC~466035000\\

466579089 & IC~2602 & 15.910 & 91\% NEB & 44\% NEB; 44\% TP$^{x}$ & TIC~466579123\\

913701079 & IC~2602 & 2.900 & 88\% NEBx2P & 84\% NEBx2P & TIC~465306570\\ 

465306494 & IC~2602 & 2.899 & 86\% NEBx2P & 85\% NEBx2P & TIC~465306570\\

378611172 & IC~2602 & 7.542 & 89\% NEBx2P & 87\% NEBx2P &  TIC~378611218\\

466422383 & IC~2602 & 33.247 & 99\% NEBx2P & 99\% NEBx2P & TIC~466422466\\

390757974 & IC~2602 & 1.705 & 54\% NEB & 54\% NEB & TIC~390758017\\

177631209 & UCL & 2.144 & 32\% EB & 21\% PEB; 21\% NEB & TIC~177631209; TIC~1167076020 \\ 

129116176 & UCL & 4.032 & 74\% EBx2P  & 91\% EBx2P & - \\

148158540 & UCL & 3.700 & 65\% EBx2P & 64\% EBx2P & -\\

\hline\hline
\end{tabular}
    \caption{List of POIs/Potential Binaries detected by \pterodactyls and vetted using \triceratops in both Gyr-old and flat prior mode. \textbf{Key:} TP = Transiting Planet around Target Star; EB = Eclipsing Binary around Target Star; EBx2P = Eclipsing Binary around Target Star with twice the orbital period compared to what was detected by \tls; NTP = Nearby Transiting Planet; NEB = Nearby Eclipsing Binary; NEBx2P = Nearby Eclipsing Binary with twice the orbital period compared to what was detected by \tls; PEB: Eclipsing Binary around Primary Star with Unresolved Bound Companion.
    $^x$See discussion in Section \ref{sub:fps} Note: The signals found in TIC~911870867 \& TIC~911870847 as well as TIC~913701079 \& TIC~465306494 have similar orbital periods with the same potential host star suggesting that they might be blended or part of the same multiple system.}
    \label{table:binaries}
\end{table*}


\section{Planet Occurrence Rates in investigated clusters}\label{sec:occ_rates}
\subsection{Detection Efficiency}
\begin{figure}[!htb]
    \centering
    \includegraphics[width=1.0\linewidth]{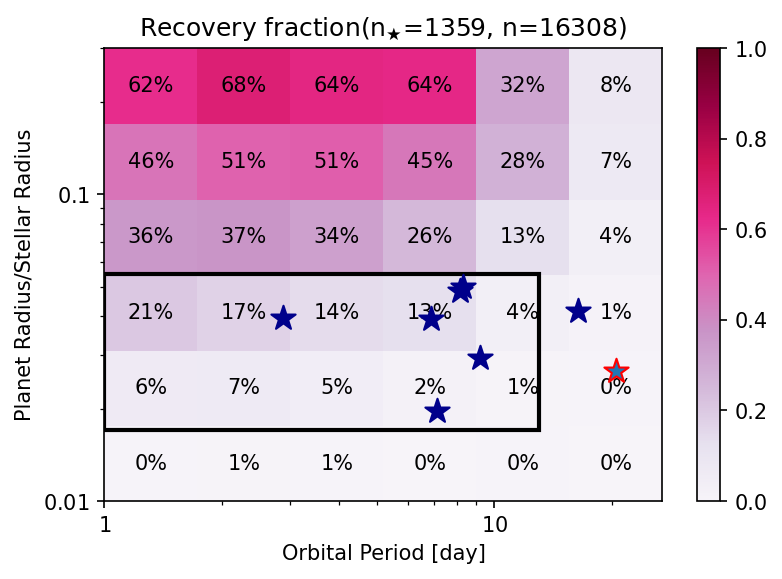}
    \caption{Overall detection efficiency of \pterodactyls while taking into account flux contamination. The confirmed planets are depicted using blue stars, except for TOI~1726\,b whose recovery required changing \pterodactyls. Darker bins represent regions of higher detection efficiency. ${n_\star}$ is the total number of stars in our sample while \textit{n} is the total number of injections done. Black box denotes the bin over which the intrinsic occurrence rates were calculated i.e., sub-Neptunes and Neptunes (0.017-0.055 \rprs\ or 1.8-6\,\Rearth, assuming a solar radius) within 12.5\,days (about half a TESS sector).}
    \label{fig:all_det_eff}
\end{figure}

An advantage of having an automated planet detection pipeline is that it enables the measurement of the survey detection efficiency which is necessary to compute intrinsic occurrence rates. The detection efficiency of our survey was calculated using injection-recovery tests on a 3$\times$4 grid in \rprs\  from 0.01 to 0.3  and in orbital period from 0.5 to 27\,days , equally spaced in both \rprs and orbital period. The injections are carried out in the \rprs\ space because most stars in our clusters currently lack stellar properties. We injected a transit signal in each bin in each light curve in our sample while taking into account the flux contribution of each star as calculated by \triceratops. We then passed these injections through \pterodactyls{} and calculated the fraction that were recovered as planets to produce a detection efficiency map per cluster as well as an average map for the entire sample. The average detection efficiency map is presented in Figure~\ref{fig:all_det_eff} and shows that the overall detection efficiency does not exceed 50\% for any bin below 0.1 \rprs (i.e., a Jupiter-sized planet assuming it orbits a Sun-like star). Interestingly, all the previously known recovered planets (denoted using blue stars in Figure~\ref{fig:all_det_eff}) are in bins with fairly low values ($<$20\%). This might be due to the intrinsic occurrence of smaller vs. larger planets being higher \citep{mulders2015planet,petigura2018california, mulders2018exoplanet}. 

\begin{figure}[!htb]
\minipage{0.46\textwidth}
  \includegraphics[width=\linewidth]{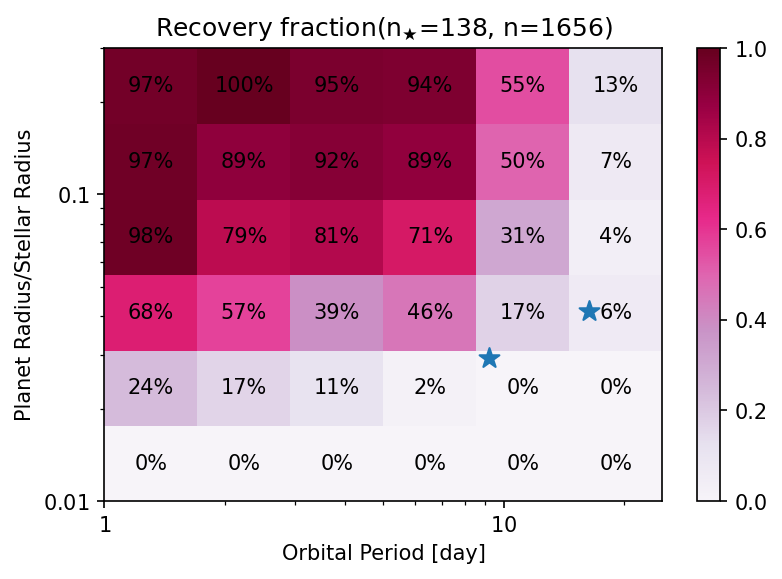}
\endminipage\hfill
\minipage{0.46\textwidth}
  \includegraphics[width=\linewidth]{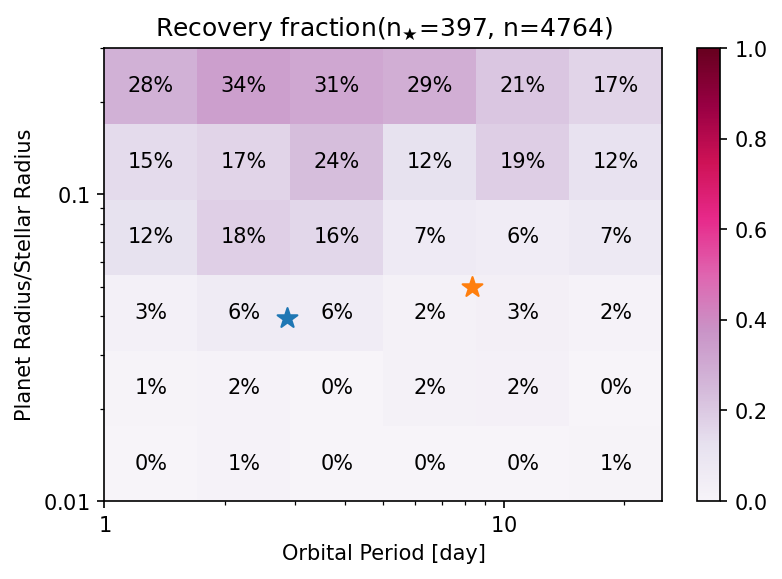}
\endminipage\hfill
\minipage{0.46\textwidth}
  \includegraphics[width=\linewidth]{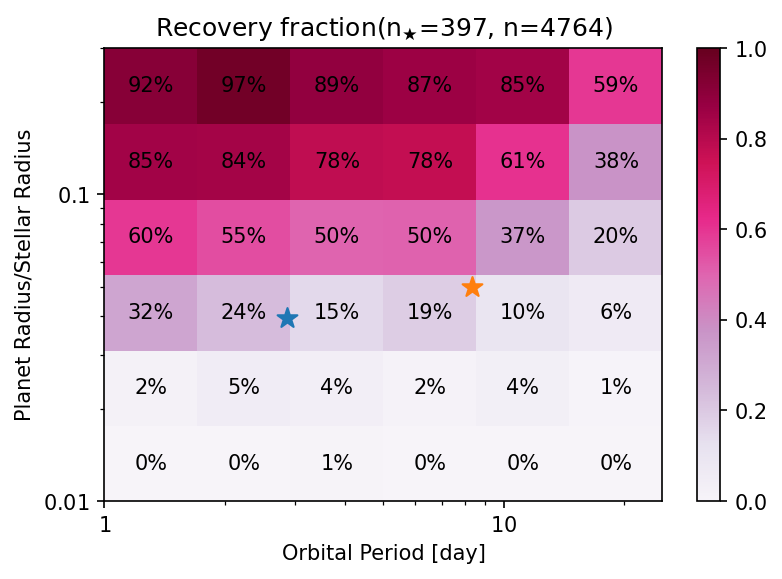}
\endminipage
\caption{A comparison of extreme detection efficiencies to highlight the effect of flux contamination: \textbf{Top: }Detection efficiency of Pisces-Eridani, a moving group with members that are spread out so flux contamination is minimal. \textbf{Middle: }Detection efficiency of IC~2602, a crowded cluster, while taking into account flux contamination while injecting planets. \textbf{Bottom: }Detection efficiency of IC~2602 while assuming no flux contamination while injecting planets.}
\label{fig:det_eff_comp}
\end{figure}

We also investigated how the detection efficiencies of each cluster contributed to the overall detection efficiency. Figure~\ref{fig:det_eff_comp} illustrates that the detection efficiency of a cluster can be quite different from the average. For example, Pisces-Eridani (top panel) has a significantly larger detection efficiency. IC~2602 (middle panel), on the other hand, has an extremely low detection efficiency. These differences might be due to age (IC~2602 is younger, hence its stars might be more variable) and/or to flux contamination due to stellar crowding, which is significantly more pronounced in IC~2602.

We further investigated the effects of flux contamination on IC~2602's detection efficiency by re-running the injection recovery tests without accounting for flux contamination. In other words, signals are injected into the light curve assuming that all of the flux comes from the target star and hence, there is no transit depth dilution. The detection efficiency map is given in the bottom panel of Figure~\ref{fig:det_eff_comp} and it is similar to that of Pisces Eridani within 10\,days and above 0.06\rprs. This test demonstrates that flux contamination significantly reduces the detectability of all planets but stellar variability in the youngest associations still plays a role for the smaller planets.

\subsection{Preliminary Calculation of Occurrence Rates}
The intrinsic occurrence rate of planets ($\eta$) can be calculated from the fraction of stars with detected planets in a survey and the survey completeness. Here, we used the inverse detection efficiency method as follows:
\begin{equation}
\eta = \frac{{n_\text{p}}}{n_\star} * \frac{1}{comp_\text{bin}}
\end{equation}
where $comp_\text{bin}$ is the survey completeness evaluated in the defined bin, $n_p$ is the number of detected planets in the bin and $n_\star$ is the number of surveyed stars. The survey completeness is computed by combining the detection efficiency and the geometric transit probability, which is given by
\begin{equation}
f_\text{geo} = \frac{R_\star}{a}
\end{equation}
Here, $R_\star$ is the stellar radius, and $a$ is the average semi-major axis which is calculated using \kepler's third law. The uncertainty on the occurrence rate was calculated from the square root of the number of detected planets in the bin, assuming Poissonian statistics.

In our analysis, we only considered sub-Neptunes and Neptunes (0.017-0.055 \rprs\ or 1.8-6\,\Rearth, assuming a solar radius) within 12.5\,days (about half a TESS sector)\footnote{Here, we use the \rprs fit by \tls and not the values from \exotic since \exotic is not part of \pterodactyls and is only used to fit for the planetary parameters once the POI is vetted by \triceratops.} because we wanted to understand the primordial population of sub-Neptunes and Neptunes before being stripped off of their atmospheres. With an average detection efficiency of 9\% and a $f_\text{geo}$ of 0.1 (at a geometric mean orbital period of 3.5\,days and assuming a Solar-type star), we compute an occurrence rate of 49$\pm$20\% for our sample of young clusters.

We then use \epos \citep{mulders2018exoplanet} to calculate the \kepler Gyr-old FGK (Sun-like) occurrence in the same bin. \epos{} is a well documented and tested Python code developed by our team\footnote{\url{http://eos-nexus.org/epos/}}, it is available on GitHub\footnote{\url{https://github.com/ GijsMulders/epos}}, and has already been used in several publications to compute occurrence rates as well as to compare planet formation models to the \kepler exoplanetary systems \citep[e.g.,][]{kopparapu2018exoplanet,Pascucci2018,fernandes2019hints,Mulders2019,Pascucci2019,mulders2020earths}. Here, we computed an occurrence rate value of 6.8$\pm$0.3\%, which is much lower than our computed value for young clusters. This could be attributed to several reasons, the major one being that our sample is heavily biased towards clusters with known planets. For example, if we increased the sample of young stars to search by a factor of six, which is about all nearby young stars within $\sim$200\,pc and assumed the same detection efficiency with no new planets, the occurrence would drop to $\sim$8\% which is the same as the \kepler value. Furthermore, since most of the stars in the sample do not have stellar radii, masses or effective temperatures, it is very likely that we are considering non-FGK stars, most likely lower mass stars \citep{henry2006solar,winters2014solar}, while calculating the occurrence rate which would make the comparison to \kepler inaccurate, where the sample is mostly Sun-like stars.  Therefore, in order to better estimate the occurrence rates of planets in young clusters, we will need to expand our search to all nearby clusters, measure the stellar parameters for those stars, and only compute the occurrence for FGK stars in our sample. However, if we continue to see this increased occurrence after we expand our sample to include all nearby clusters and moving groups, it could imply that there is an excess of larger planets at young ages possibly because their atmospheres have not yet been stripped due to photoevaporation and/or core-powered mass loss.


\section{Summary and Future Work}\label{sec:summary}
In this paper, we present \pterodactyls $-$ a Python-based pipeline that is built using publicly available codes, and is specifically designed to extract, detrend, search for, and vet transiting exoplanets in young clusters using TESS' Primary Mission 30-min cadence FFIs. Here, we only consider 5 specific young clusters and moving groups: Tucana-Horologium Association, IC~2602, Upper Centaurus Lupus, Ursa Major, and Pisces Eridani. These clusters were chosen solely due to the presence of eight confirmed planets and two candidates which were used to validate \pterodactyls. Using \pterodactyls,

\begin{itemize}[leftmargin=*]
    \item We were able to recover eight out of ten known transiting planets/candidates. For the two that we were not able to recover (both initially discovered using 2-min cadence data), one planet was too small to detect (PiEri; TOI~451\,b) and the other is a long period, single transit planet (UCL; HIP~67522\,c), which is beyond the capabilities of the pipeline.
    \item We also detected 15 transiting signals, all of which were found to be either eclipsing binaries or nearby eclipsing binaries when their phase-folded light curves were vetted by \triceratops.
    \item While taking into account the stellar flux contribution, we conducted injection-recovery tests in order to measure the detection efficiency of the \pterodactyls. We found that the overall detection efficiency does not exceed 50\% for any bin below 0.1 \rprs (i.e., a Jupiter-sized planet orbiting a Sun-like star) and that all the recovered planets are in bins with fairly low values ($<$20\%), hinting that even in young clusters smaller planets are more frequent than giant planets.
    \item For our biased sample, we computed a planet occurrence rate of 49$\pm$20\% which is significantly higher than \kepler's Gyr-old occurrence rates of 6.8$\pm$0.3\% for sub-Neptunes and Neptunes (0.017-0.055 \rprs\ or 1.8-6\,\Rearth, assuming a Solar radius) within 12.5\,days (about half a TESS sector).
\end{itemize}

Since most of these planets were initially detected in 2-min cadence data, our work shows that 30-min cadence data can be used to detect young transiting planets and thus has the potential for more planet discoveries. The planet occurrence rate we calculated for our sample is higher than that for \kepler's Gyr old stars. However, we realize that this value is heavily biased since we only consider clusters with confirmed/candidate planets and can be further improved upon by using \pterodactyls to search for young ($< 1$\,Gyr) transiting planets in a larger sample of nearby clusters and moving groups ($<$200\,pc) using TESS Primary and Extended Mission FFIs. We also plan on uniformly characterizing all the stars in our sample, so as to only consider FGK stars in future studies. With this young population of transiting, short-period planets, we hope to establish how the radius distribution of transiting exoplanets evolved over time, and therefore provide observational constraints on the mass loss mechanisms of planetary atmospheres.


\software{\texttt{NumPy} \citep{numpy}, \texttt{SciPy} \citep{scipy}, \texttt{Matplotlib} \citep{pyplot}, \eleanor{} \citep{feinstein2019eleanor}, \wotan{} \citep{hippke2019wotan}, \texttt{transitleastsquares} \citep{hippke2019optimized}, \texttt{vetting} \citep{hedges2021vetting}, \triceratops{} \citep{giacalone2020vetting}, \texttt{EDI-Vetter Unplugged} \citep{zink2019edivetter}, \exotic \citep{zellem2020utilizing}, \epos \citep{Mulders2018}}


\section{Acknowledgements}
R.B.F. would like to thank the following individuals for their expertise, assistance and, invaluable insights throughout the testing of \pterodactyls: Adina D. Feinstein and Benjamin T. Montet (\eleanor), Michael Hippke (\wotan and \tls), Robert T. Zellem (\exotic), Christina Hedges (\texttt{vetting}, centroid test), and the Scaling \ktwo team i.e., Jessie L. Christiansen, Sakhee Bhure and Britt Duffy Adkins. G.D.M. acknowledges support from ANID --- Millennium Science Initiative --- ICN12\_009. This paper includes data collected by the TESS mission. Funding for the TESS mission is provided by the NASA's Science Mission Directorate. This material is based upon work supported by the National Aeronautics and Space Administration (NASA) under Agreement No. NNX15AD94G for the program Earths in Other Solar Systems, under Agreement No. 80NSSC21K0593 for the program “Alien Earths”, and Grant No. 80NSSC20K0446 issued through the Astrophysics Data Analysis Program (ADAP). The results reported herein benefited from collaborations and/or information exchange within NASA’s Nexus for Exoplanet System Science (NExSS) research coordination network sponsored by NASA’s Science Mission Directorate. This publication makes use of data products from Exoplanet Watch, a citizen science project managed by NASA’s Jet Propulsion Laboratory on behalf of NASA’s Universe of Learning. This work is supported by NASA under award number NNX16AC65A to the Space Telescope Science Institute.
\clearpage

\bibliographystyle{apj}
\bibliography{main}
\clearpage
\appendix
\section{Comparison of Spline-Based Detrending}\label{sec:detrend_test}
In order to find the best detrending routine for our sample, we followed \citet{hippke2019wotan} which used injection-recovery tests and found that for highly variable stars, such as those found in young cluster, spline-based methods maximized the fraction of recovered injected planets. We specifically tested the following three splines available via \wotan: robust spline with iterative sigma clipping, Huber estimator spline \citet{huber1981robust}, and a penalized spline with iterative sigma clipping \citet{eilers1996flexible}. Here, we demonstrate how each of these splines detrend the light curve of HIP~67522 which is in UCL and also has two planets around it. As can be seen below, the penalized spline does the best job of detrending the light curve and has the lowest RMS.

\begin{figure*}[!htb]
   \includegraphics[width=\textwidth]{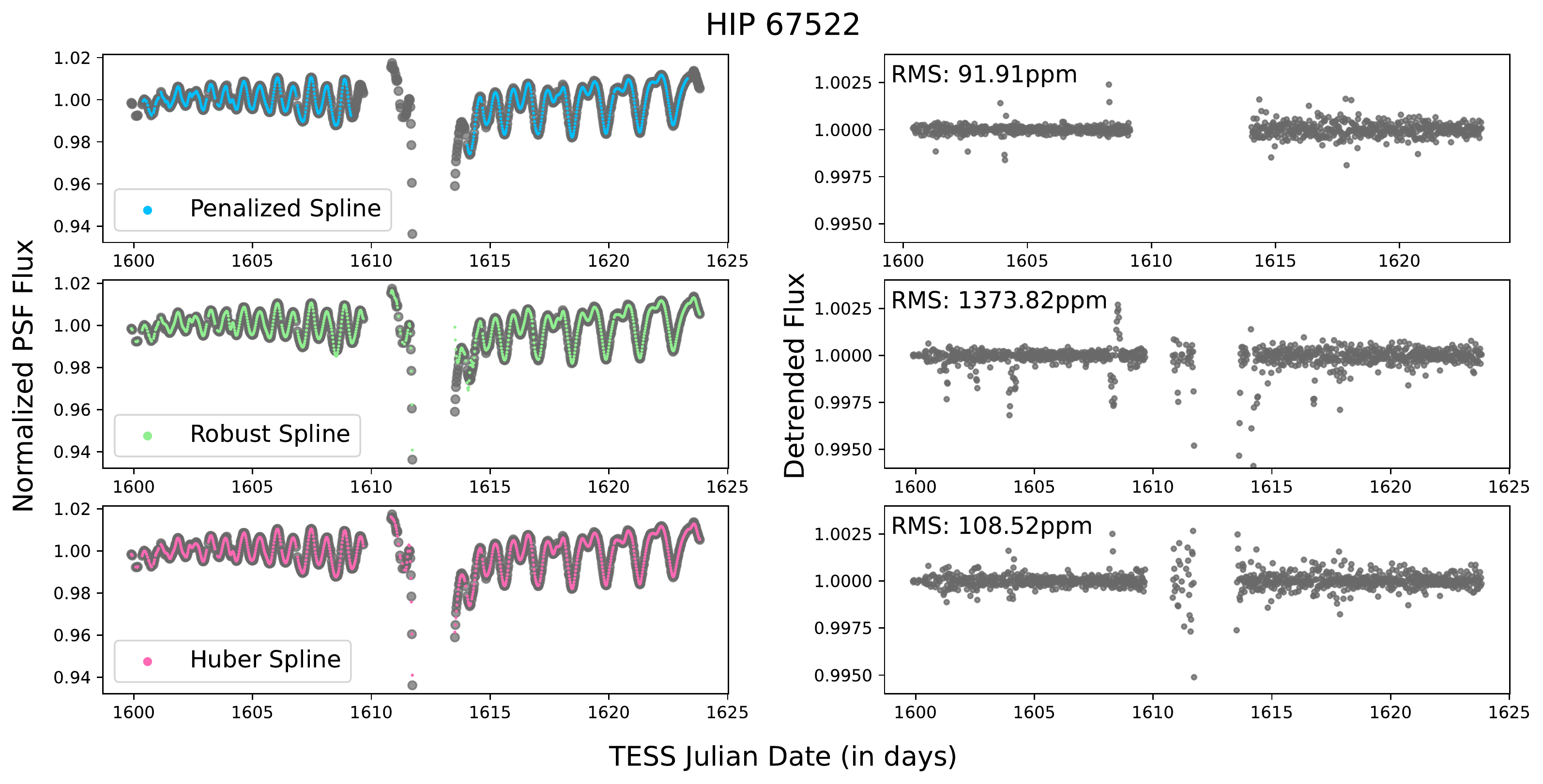}
\caption{A demonstration of how different splines detrend the light curve of HIP~67522. As can be seen by the low RMS, the penalized spline does a superior job of matching the stellar variability trend and detrending the light curve.}
\label{fig:detrend}
\end{figure*}

\clearpage
\section{Recovery of Known Planets}\label{sec:recovery}
The following plots depict the recovery of the known young transiting planets using \pterodactyls: 

\begin{figure}[!htb]
\minipage{0.5\textwidth}
  \includegraphics[width=\linewidth]{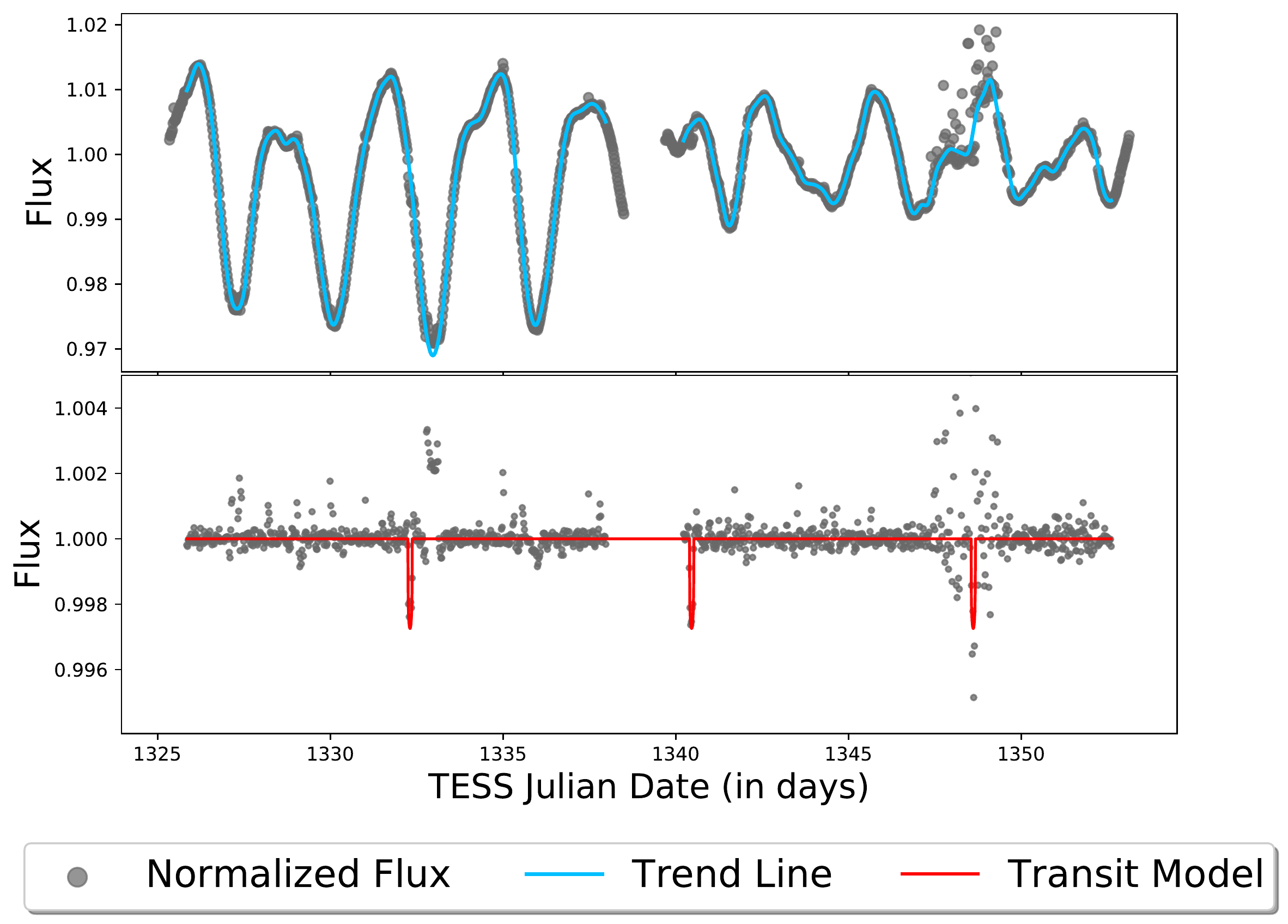}
\endminipage\hfill
\minipage{0.5\textwidth}
  \includegraphics[width=\linewidth]{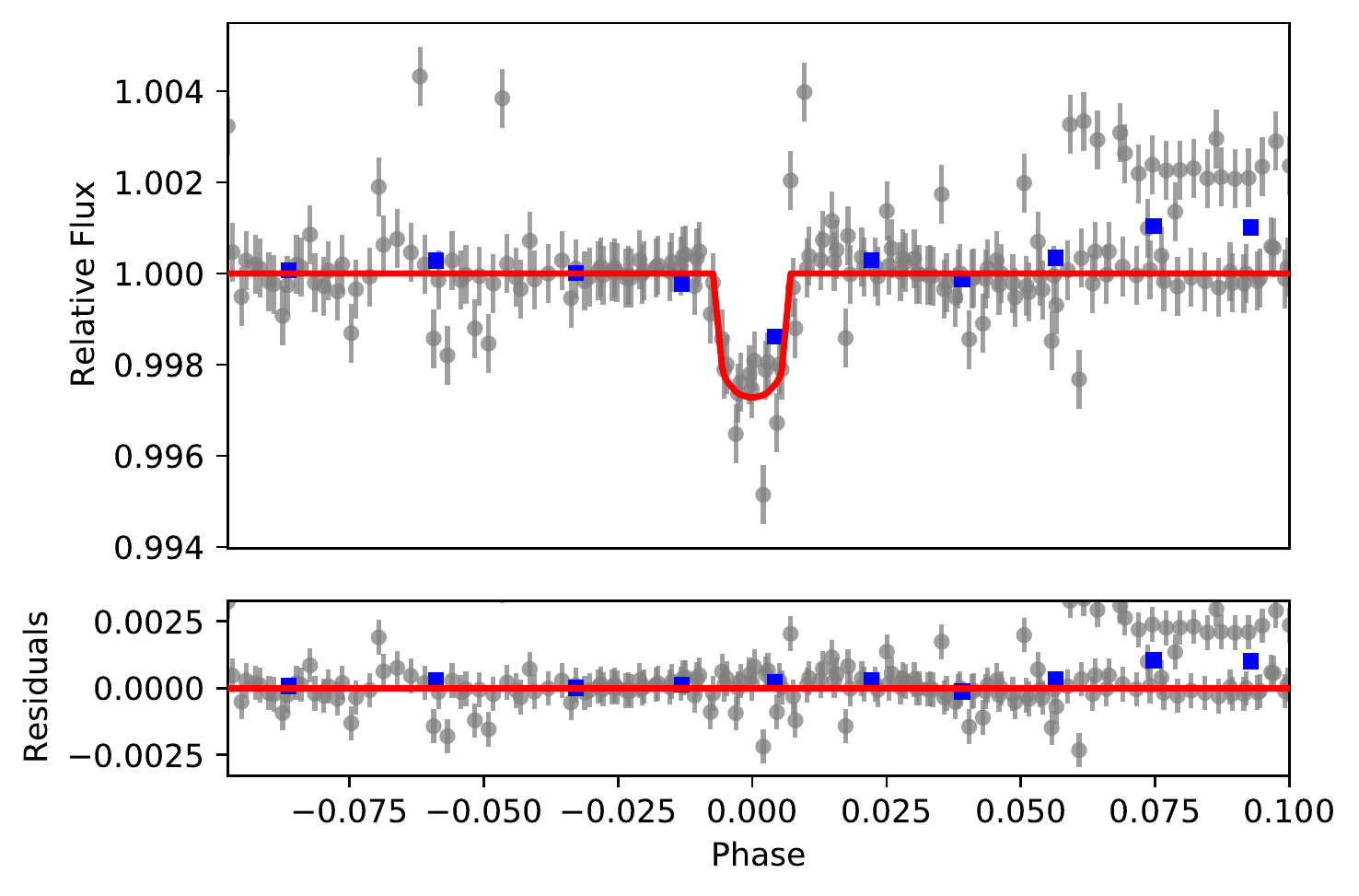}
\endminipage
\caption{\textbf{Left: }Progression of the DS Tuc A light curve through \pterodactyls{}. Upper panel: PSF flux (in dark gray) and penalized spline trend (in blue). Lower panel: Detrended light curve (PSF flux divided by the trend). The multi-transit signature (in red) can be clearly seen. \textbf{Right: } Upper panel shows Phase-folded light curve of DS Tuc A\,b (gray data points) along with the best fit model using \exotic (in red). Lower panel shows the residuals in parts per million.}
\label{fig:dstuc}
\end{figure}

\begin{figure}[!htb]
\minipage{0.5\textwidth}
  \includegraphics[width=\linewidth]{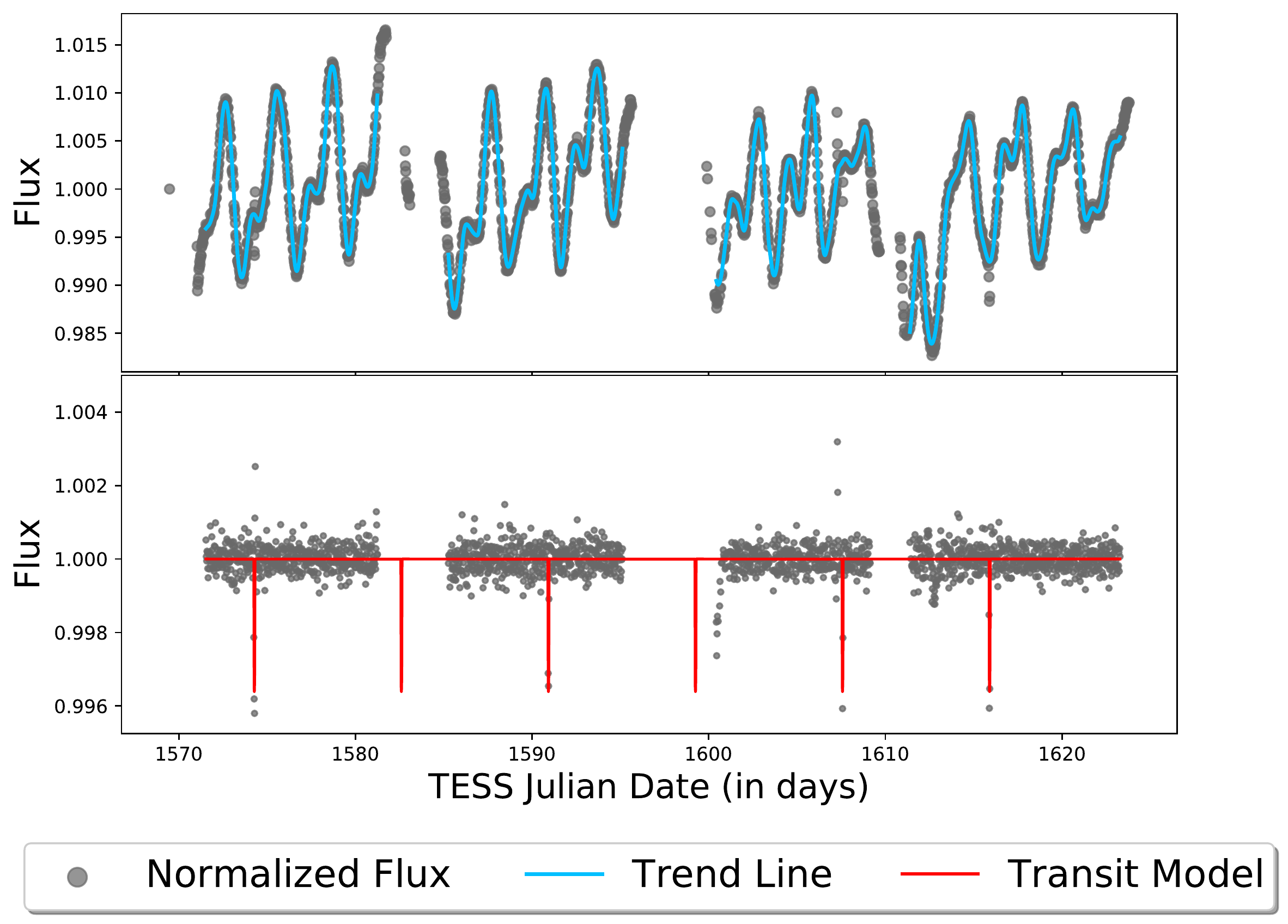}
\endminipage\hfill
\minipage{0.5\textwidth}
  \includegraphics[width=\linewidth]{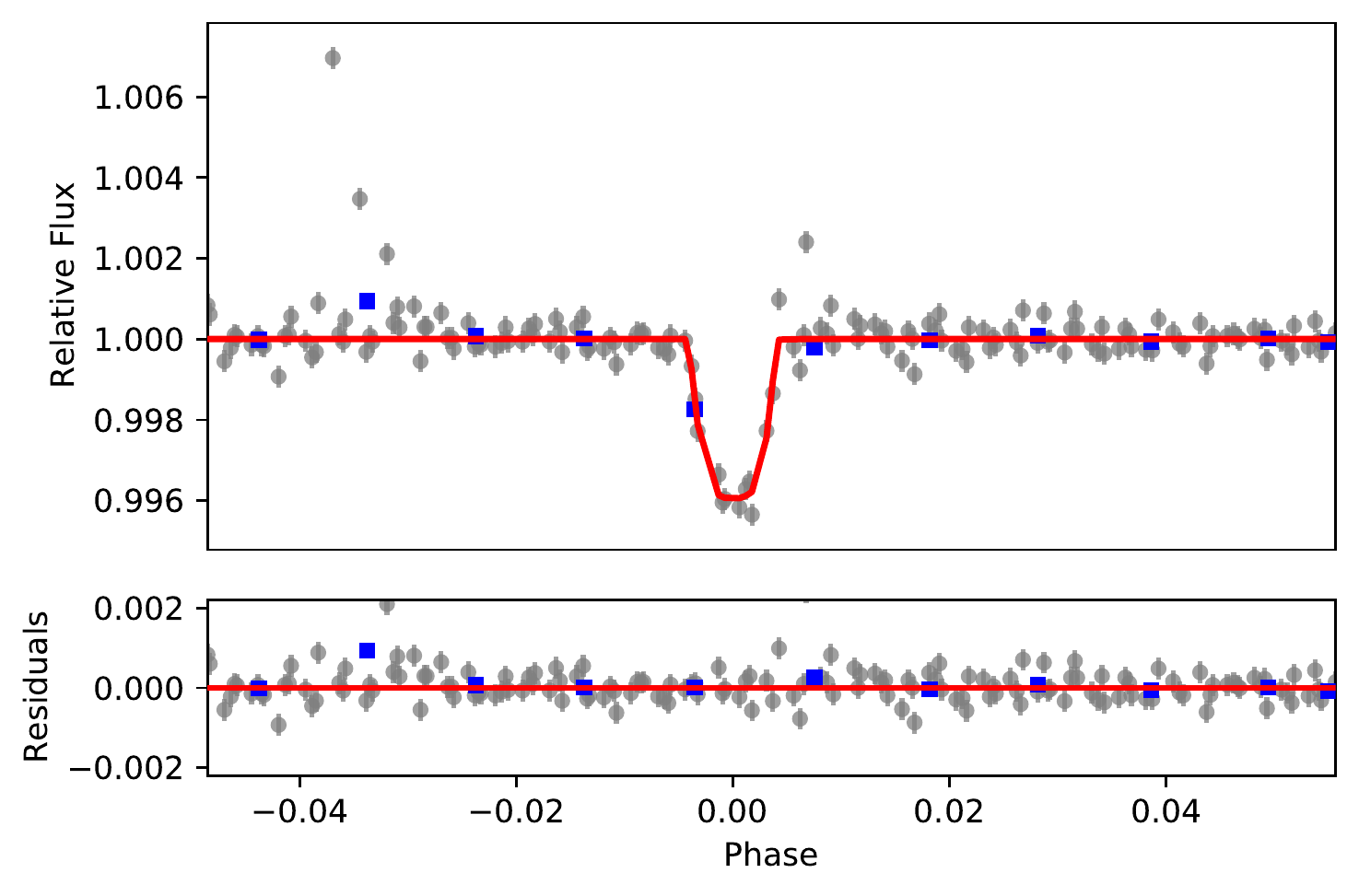}
\endminipage
\caption{\textbf{Left:} Recovery of TIC~460205581\,b. Upper panel: PSF flux (in dark gray) and penalized spline trend (in blue). Lower panel: Detrended light curve (PSF flux divided by the trend). The multi-transit signature (in red) can be clearly seen. \textbf{Right:} Upper panel shows Phase-folded light curve of TIC~460205581 b (gray data points) along with the best fit model using \exotic (in red). Lower panel shows the residuals in parts per million.}
\label{fig:ic2602_5581}
\end{figure}

\begin{figure}[!htb]
\minipage{0.5\textwidth}
  \includegraphics[width=\linewidth]{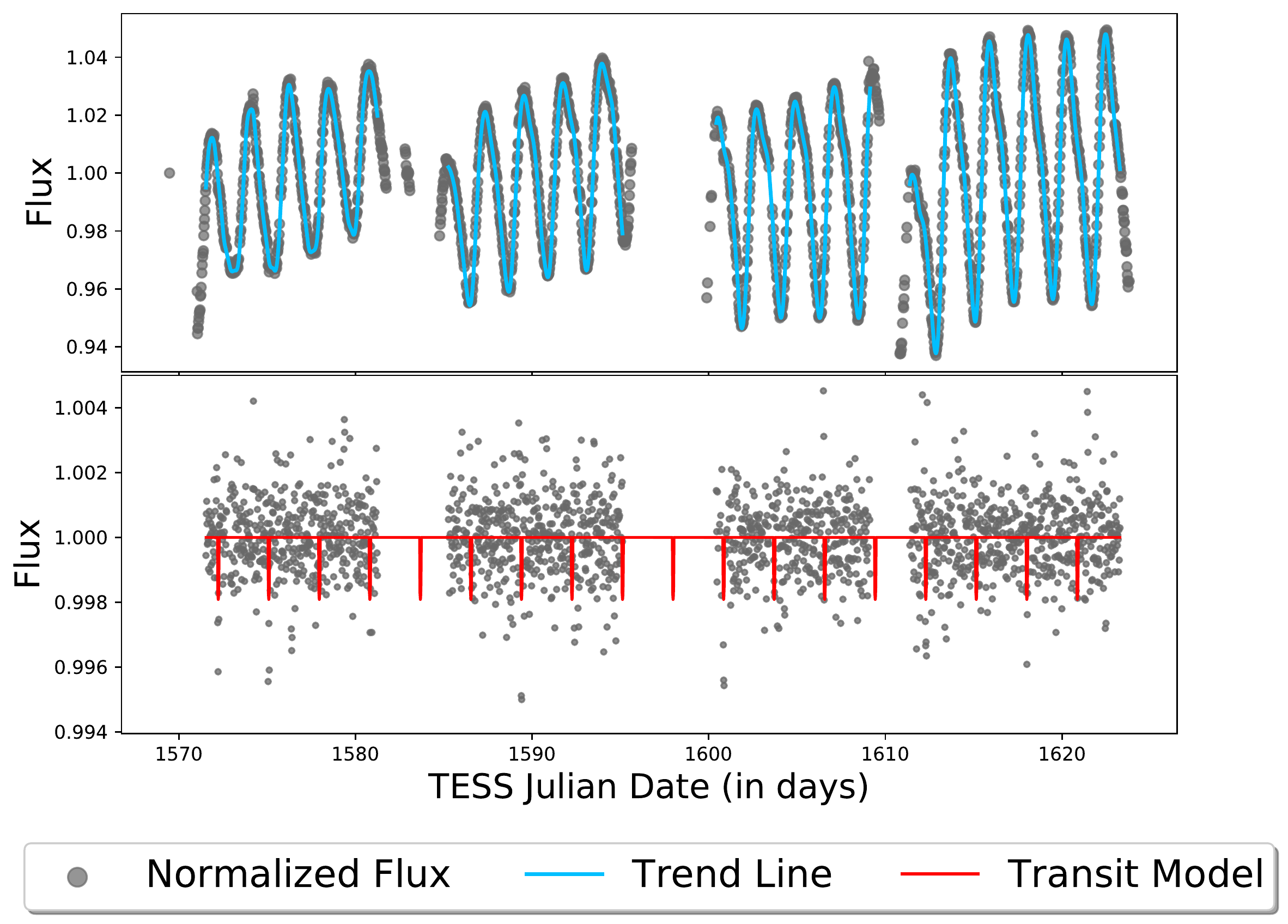}
\endminipage\hfill
\minipage{0.5\textwidth}
  \includegraphics[width=\linewidth]{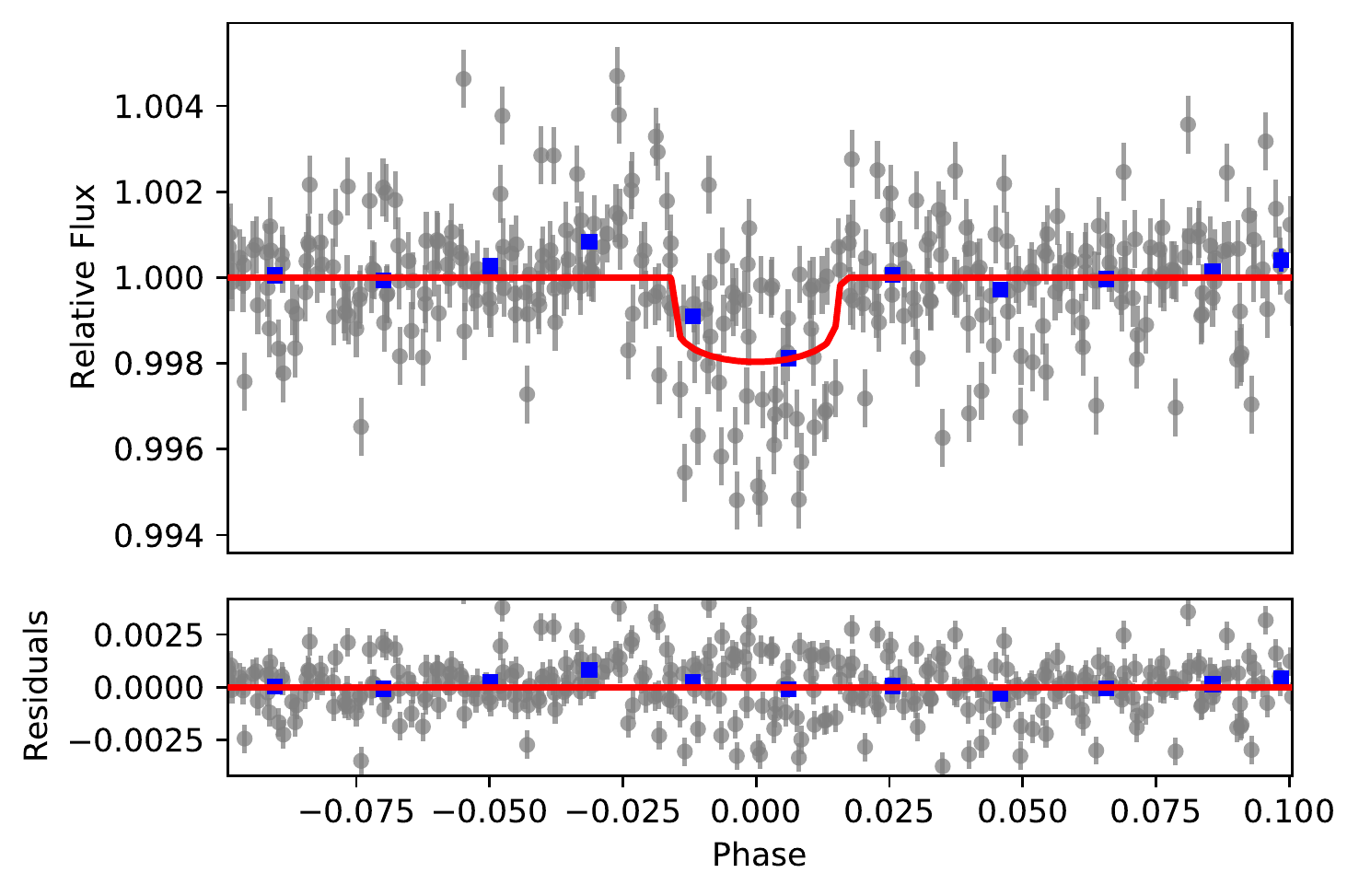}
\endminipage
\caption{\textbf{Left:} Recovery of TOI~837\,b. Upper panel: PSF flux (in dark gray) and penalized spline trend (in blue). Lower panel: Detrended light curve (PSF flux divided by the trend). The multi-transit signature (in red) can be clearly seen. \textbf{Right:} Upper panel shows Phase-folded light curve of TOI~837\,b (gray data points) along with the best fit model using \exotic (in red). Lower panel shows the residuals in parts per million.}
\label{fig:ic2602_0389}
\end{figure}

\begin{figure}[!htb]
\minipage{0.5\textwidth}
  \includegraphics[width=\linewidth]{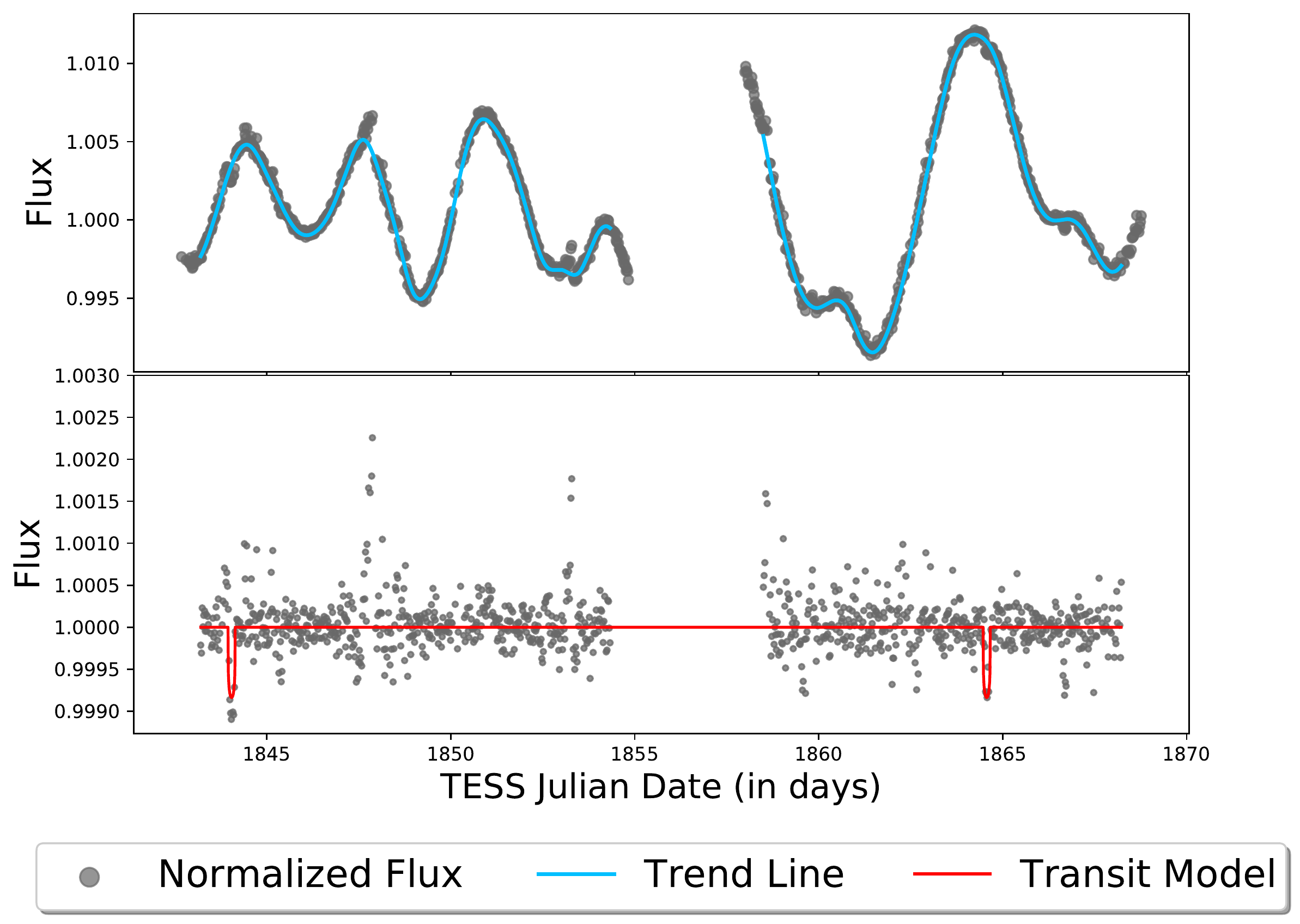}
\endminipage\hfill
\minipage{0.5\textwidth}
  \includegraphics[width=\linewidth]{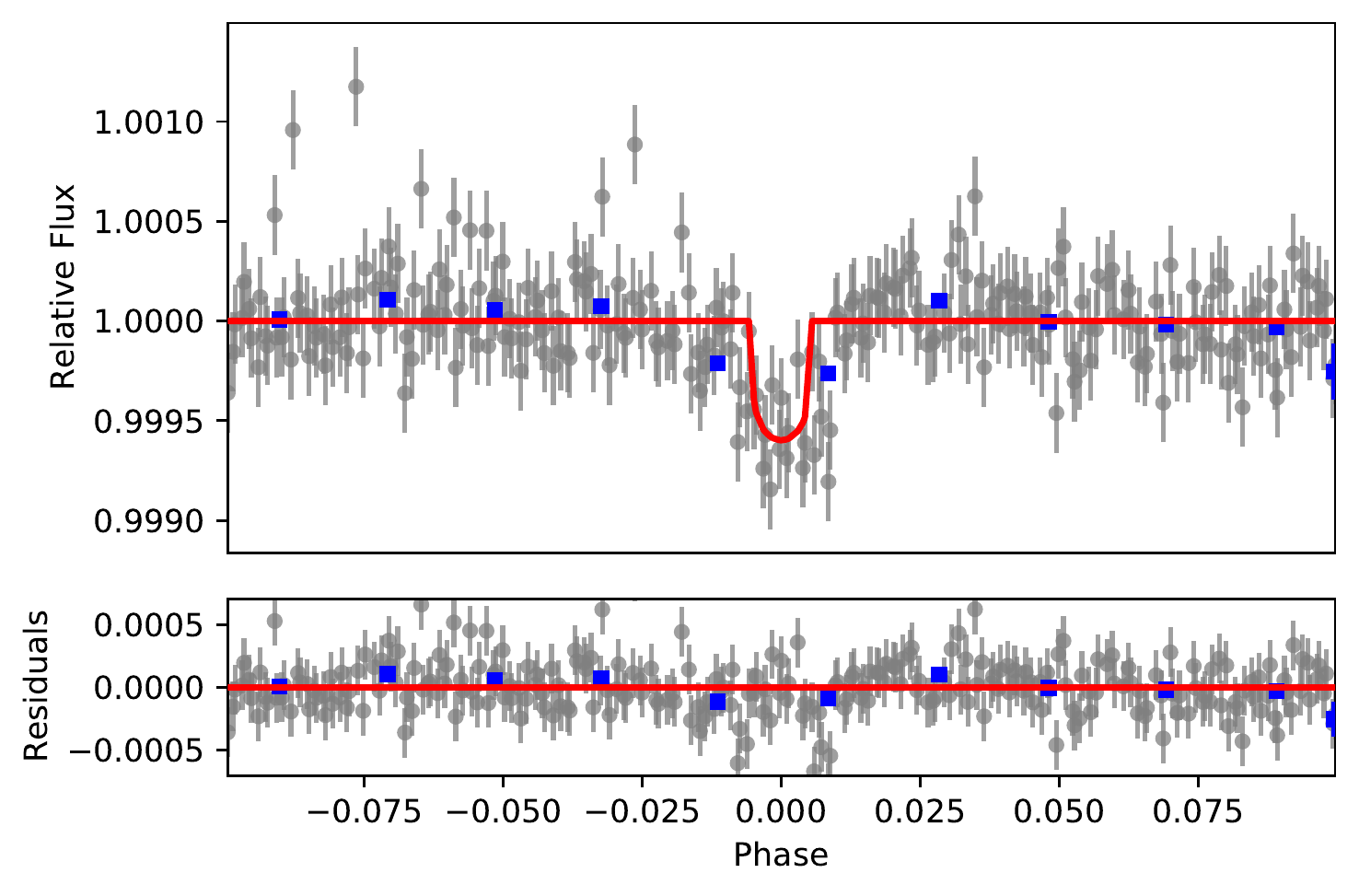}
\endminipage
\caption{\textbf{Left:} Recovery of TOI~1726\,c. Upper panel: PSF flux (in dark gray) and penalized spline trend (in blue). Lower panel: Detrended light curve (PSF flux divided by the trend). The multi-transit signature (in red) can be clearly seen. \textbf{Right:} Upper panel shows Phase-folded light curve of TOI~1726\,c (gray data points) along with the best fit model using \exotic (in red). Lower panel shows the residuals in parts per million.}
\label{fig:uma_c}
\end{figure}

\begin{figure}[!htb]
\minipage{0.5\textwidth}
  \includegraphics[width=\linewidth]{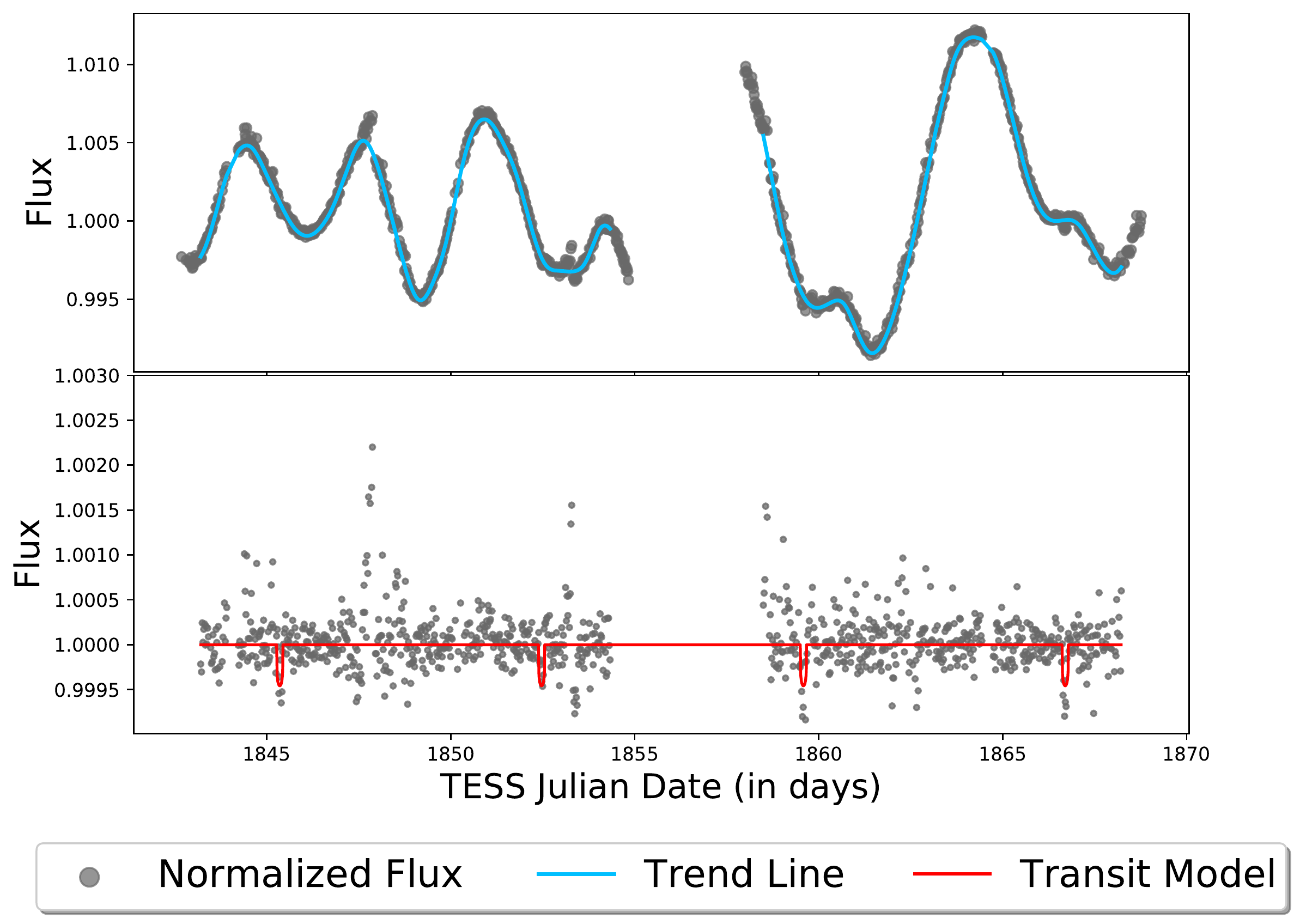}
\endminipage\hfill
\minipage{0.5\textwidth}
  \includegraphics[width=\linewidth]{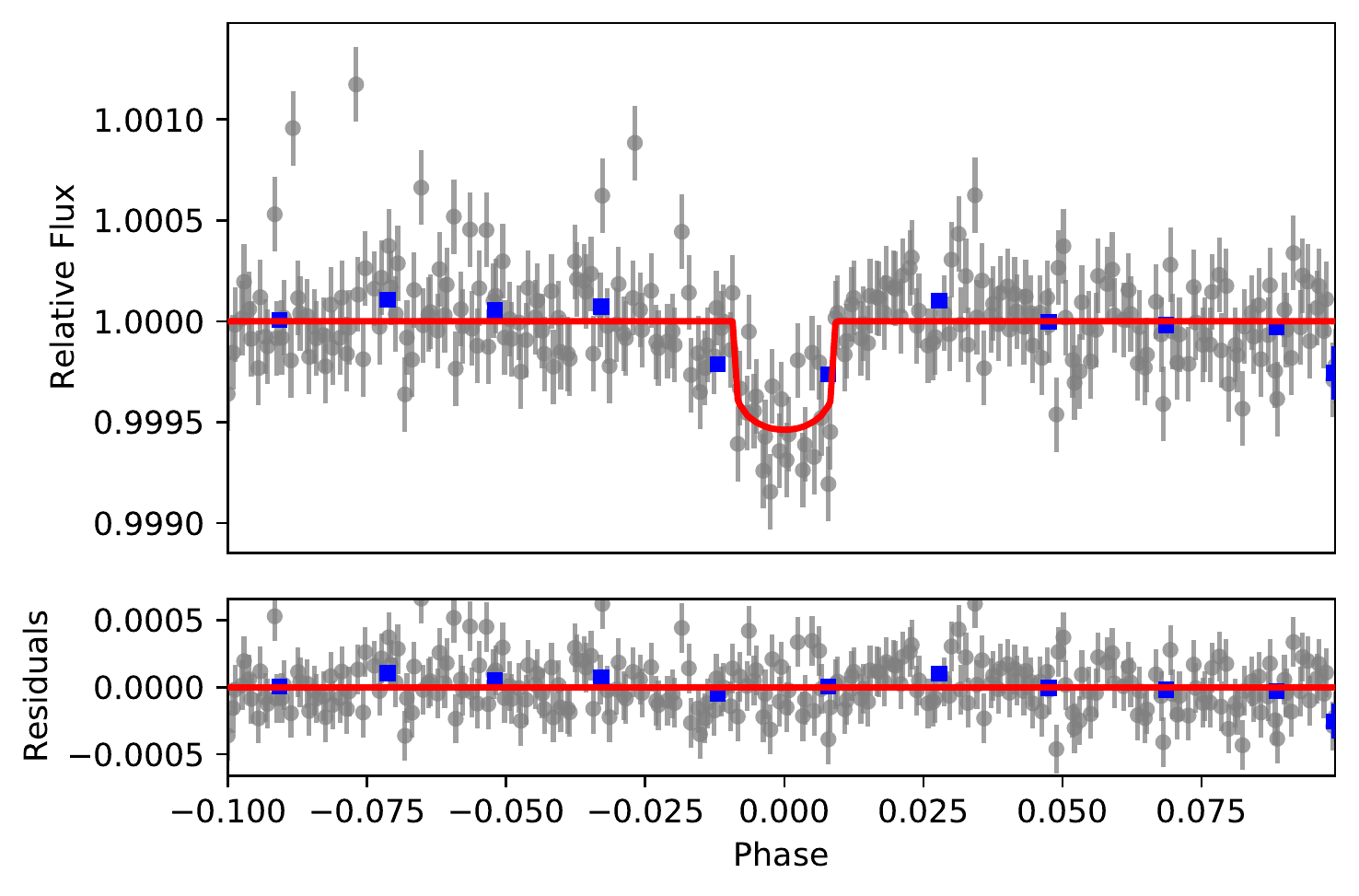}
\endminipage
\caption{\textbf{Left:} Recovery of TOI~1726\,b. Upper panel: PSF flux (in dark gray) and penalized spline trend (in blue). Lower panel: Detrended light curve (PSF flux divided by the trend). The multi-transit signature (in red) can be clearly seen. \textbf{Right:} Upper panel shows Phase-folded light curve of TOI~1726\,b (gray data points) along with the best fit model using \exotic (in red). Lower panel shows the residuals in parts per million.}
\label{fig:uma_b}
\end{figure}

\begin{figure}[!htb]
\minipage{0.5\textwidth}
  \includegraphics[width=\linewidth]{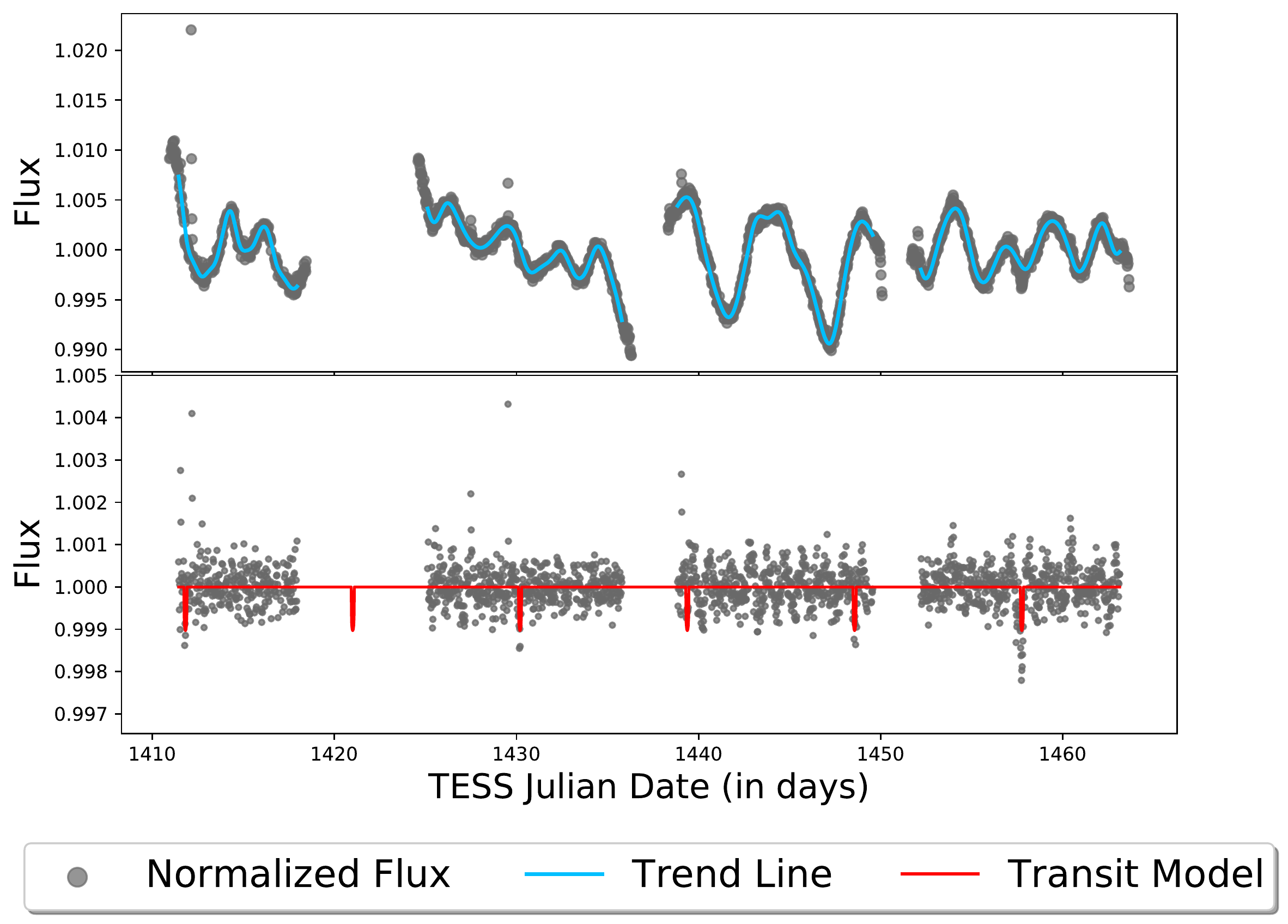}
\endminipage\hfill
\minipage{0.5\textwidth}
  \includegraphics[width=\linewidth]{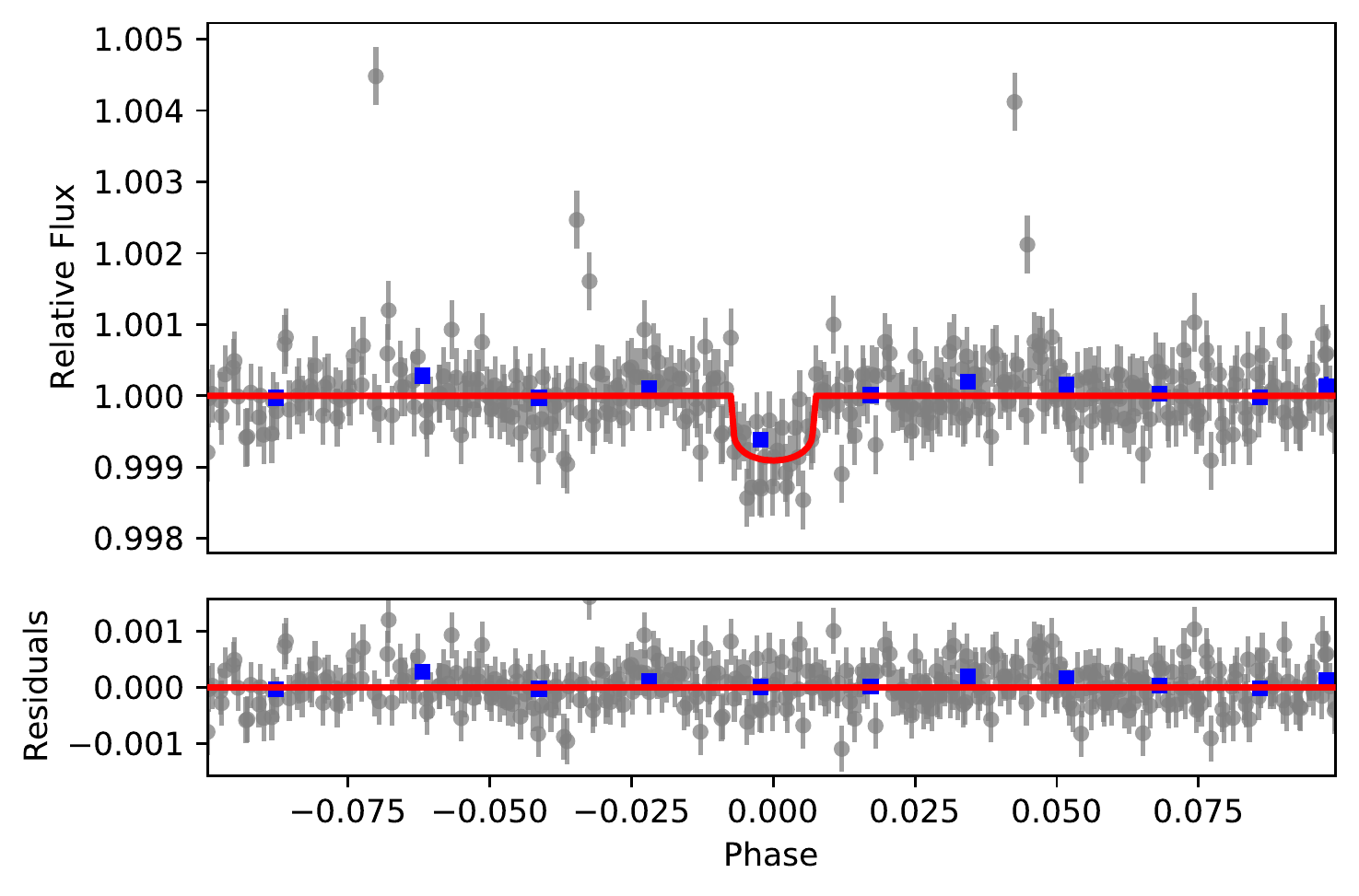}
\endminipage
\caption{\textbf{Left:} Recovery of TOI~451\,c. Upper panel: PSF flux (in dark gray) and penalized spline trend (in blue). Lower panel: Detrended light curve (PSF flux divided by the trend). The multi-transit signature (in red) can be clearly seen. \textbf{Right:} Upper panel shows Phase-folded light curve of TOI~451\,c (gray data points) along with the best fit model using \exotic (in red). Lower panel shows the residuals in parts per million.}
\label{fig:pieri_c}
\end{figure}

\clearpage

\section{Centroid Test}\label{sec:centroid}
\subsection{Recovered Known Transiting Planets}
The following plots show the results of the centroid tests for the recovered, known, young transiting planets using \pterodactyls. We find that, in all cases, no significant offset was detected i.e., the transiting signal was indeed coming from the target star.

\begin{figure}[!htb]
\minipage{0.5\textwidth}
  \includegraphics[width=\linewidth]{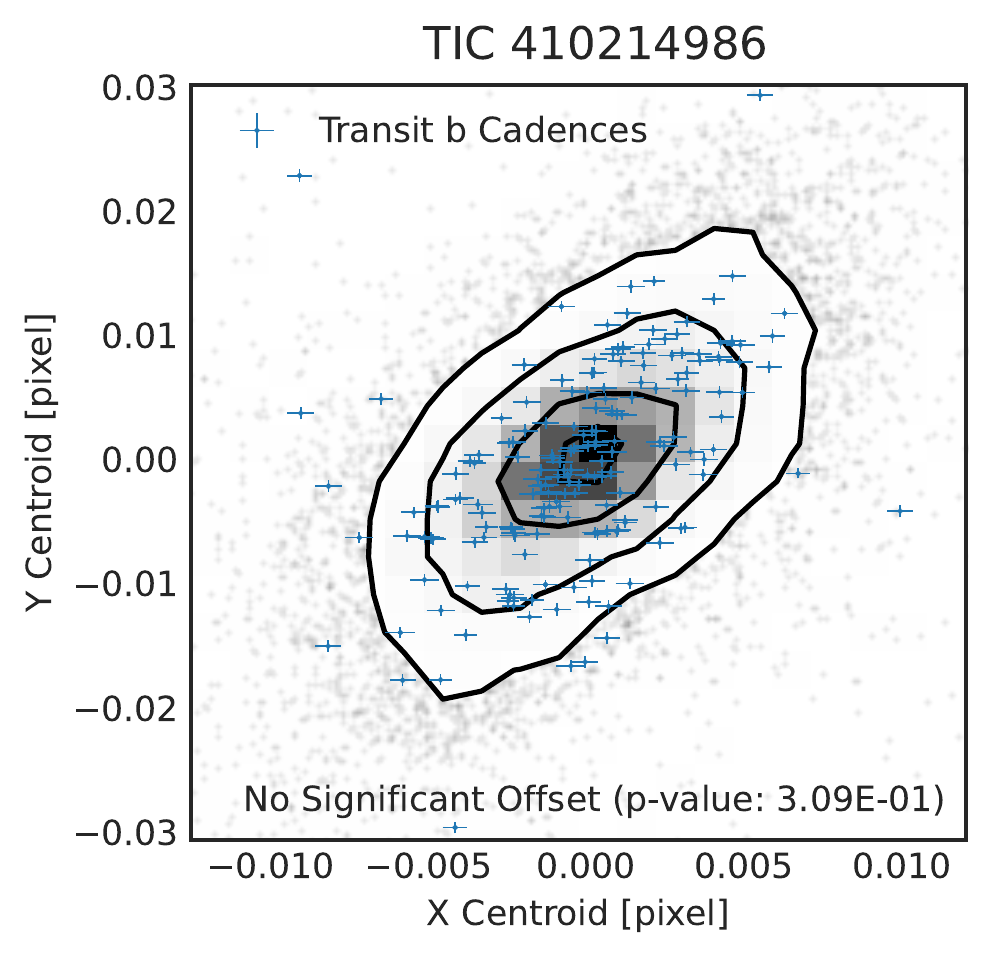}
\endminipage\hfill
\minipage{0.5\textwidth}
  \includegraphics[width=\linewidth]{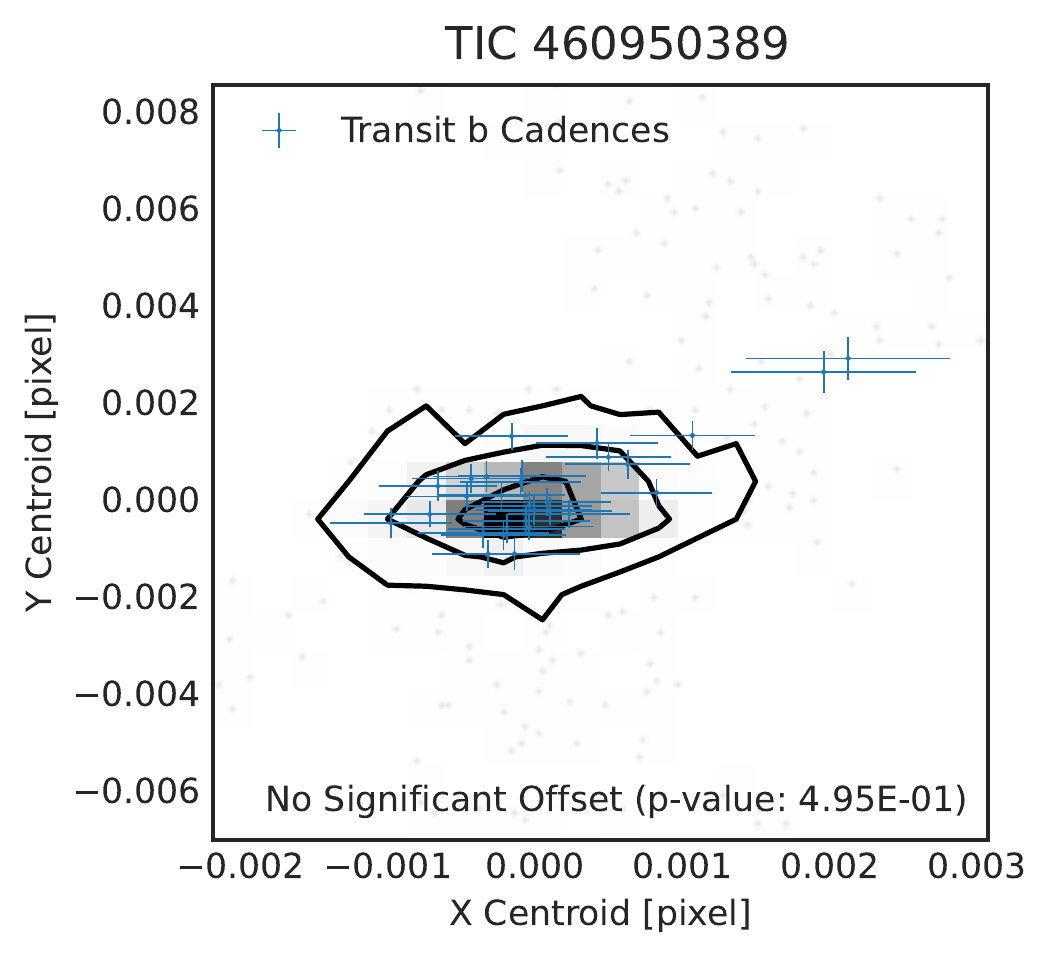}
\endminipage\hfill
\minipage{0.5\textwidth}
  \includegraphics[width=\linewidth]{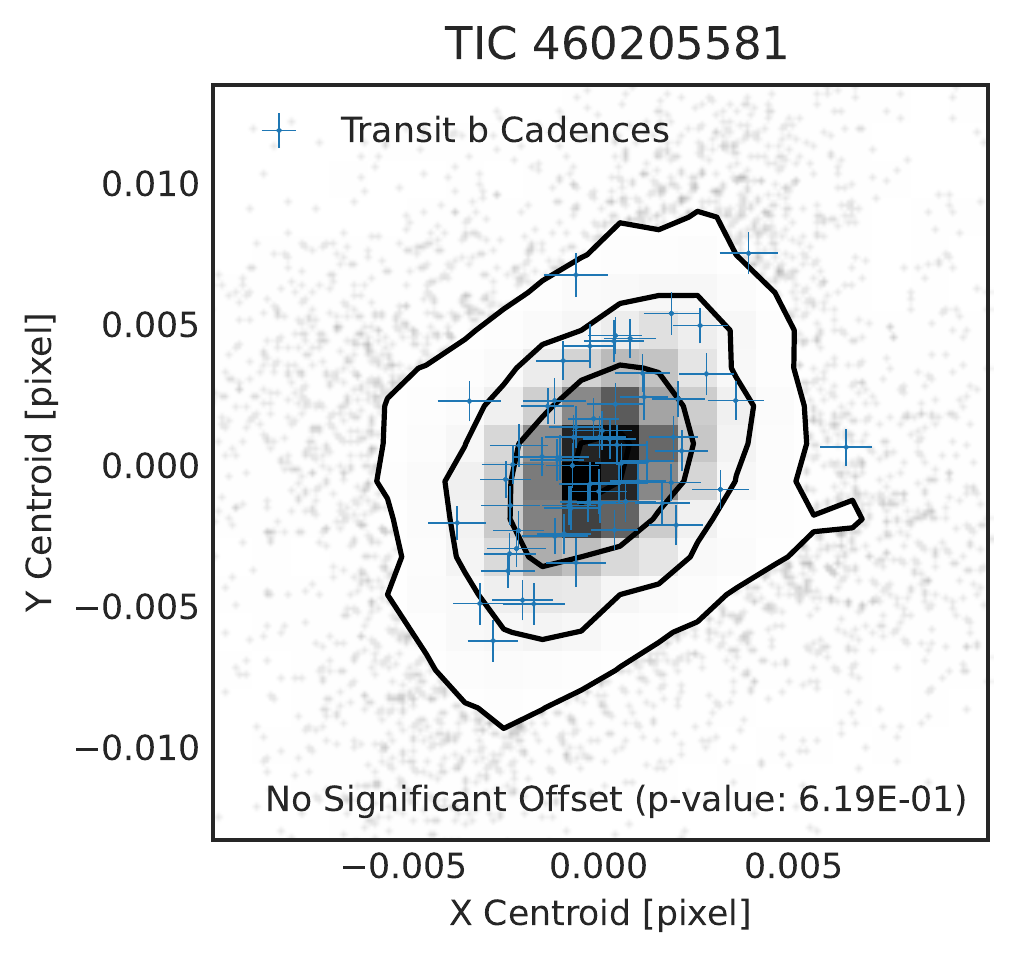}
\endminipage\hfill
\minipage{0.5\textwidth}
  \includegraphics[width=\linewidth]{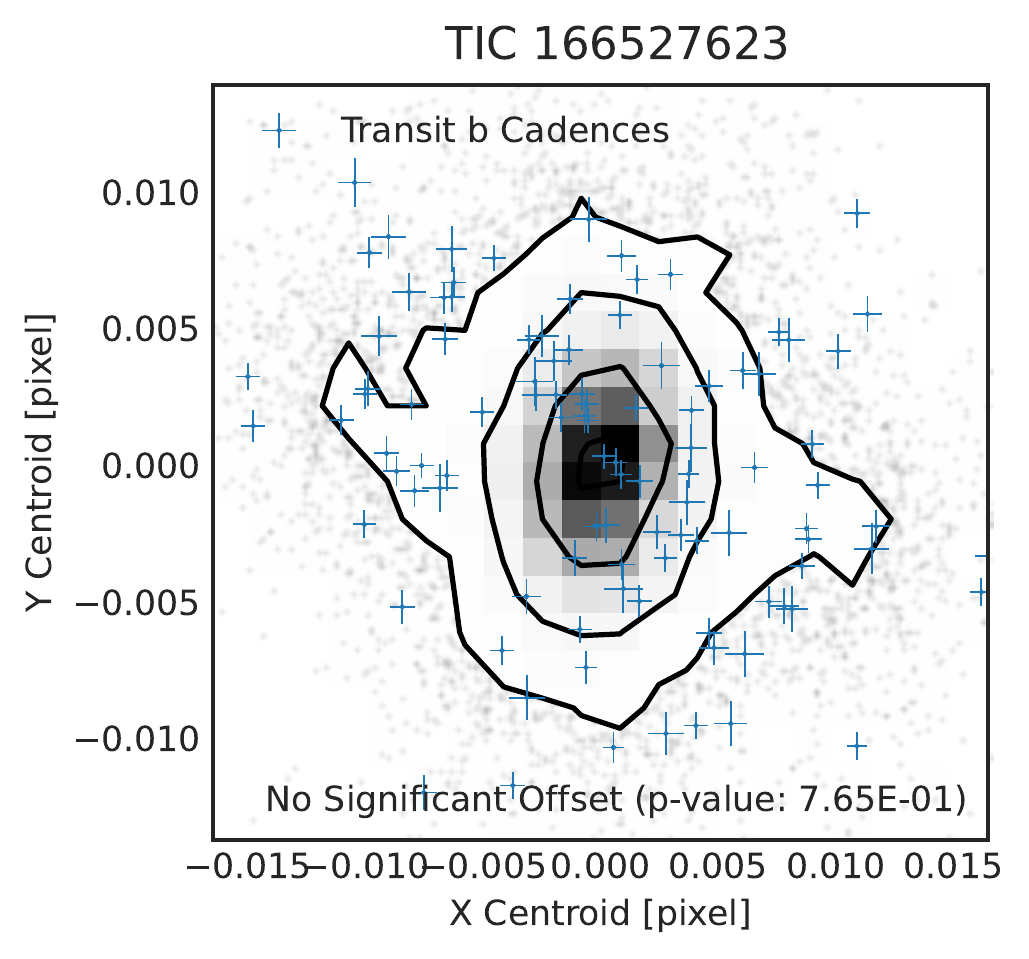}
\endminipage
\caption{Centroid tests for single planet systems}
\label{fig:centroid_1}
\end{figure}

\begin{figure}[!htb]
\minipage{\textwidth}
  \includegraphics[width=\linewidth]{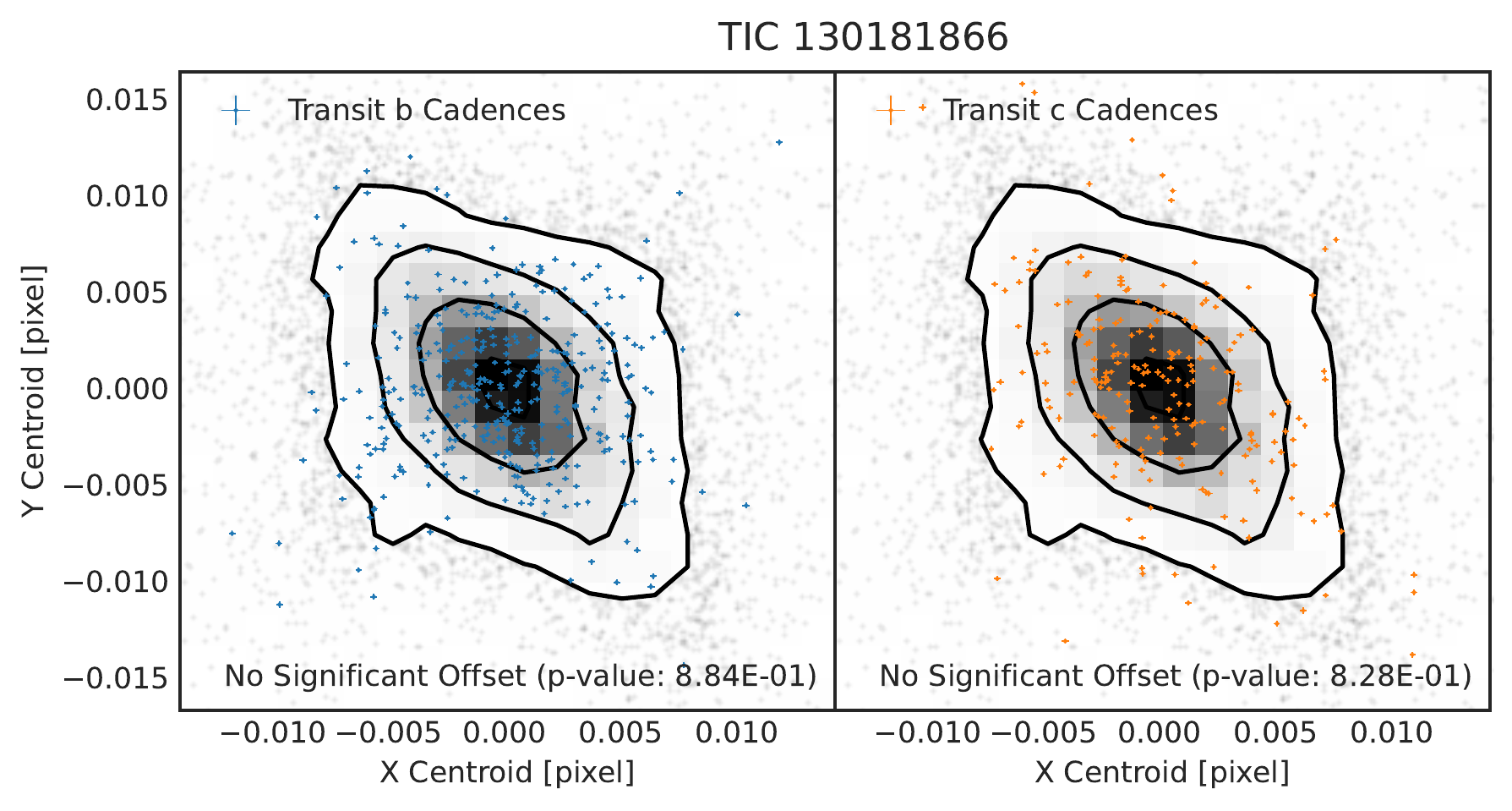}
\endminipage\hfill
\minipage{\textwidth}
  \includegraphics[width=\linewidth]{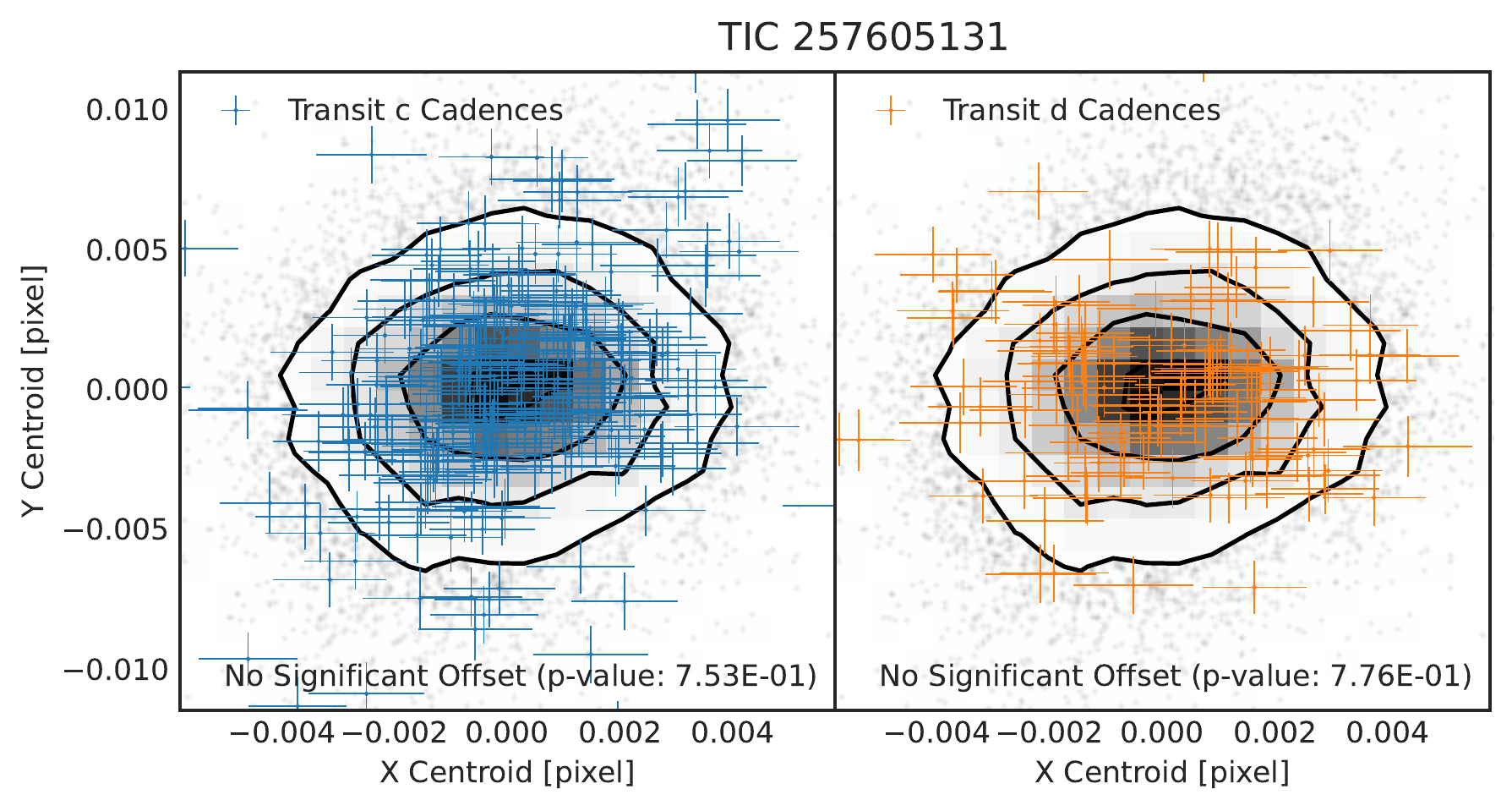}
\endminipage
\caption{Centroid tests for multi-planet systems}
\label{fig:centroid_2}
\end{figure}

\clearpage
\subsection{POIs/Potential Binaries}
\citet{hedges2021vetting} uses the TESS Target Pixel File (TPF; and the aperture information therein) in order to conduct the centroid test. However, TPFs were only available for three of the potential binaries (all of the ones in UCL; shown below) in Table~\ref{table:binaries} but not for the 12 other candidates (IC2602). As such, we were only able to use this test when the TPF was available. As can be seen below in two of the cases, we see no significant offsets. However, in the case of TIC~177631209, an offset is detected which is consistent with \triceratops' Flat prior prediction i.e., the transiting signal is actually coming from a nearby/background star.

\begin{figure}[!htb]
\minipage{0.5\textwidth}
  \includegraphics[width=\linewidth]{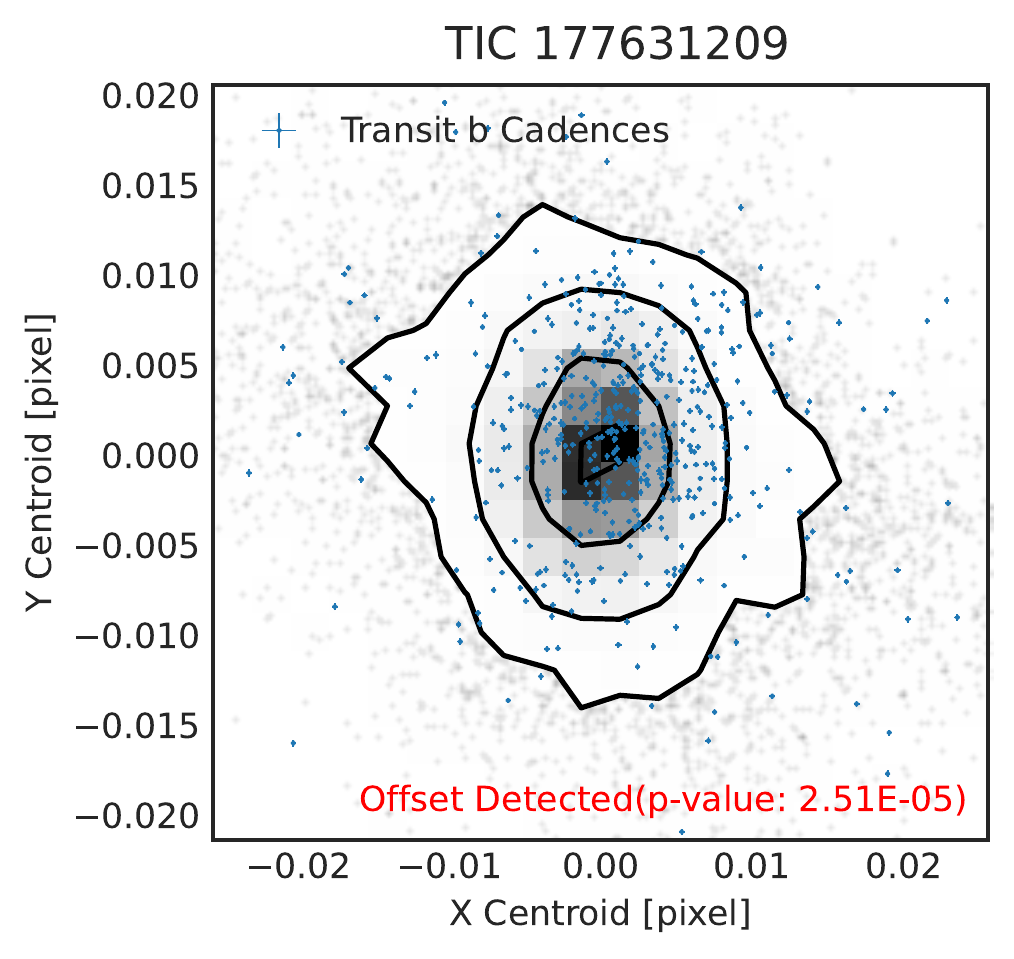}
\endminipage\hfill
\minipage{0.5\textwidth}
  \includegraphics[width=\linewidth]{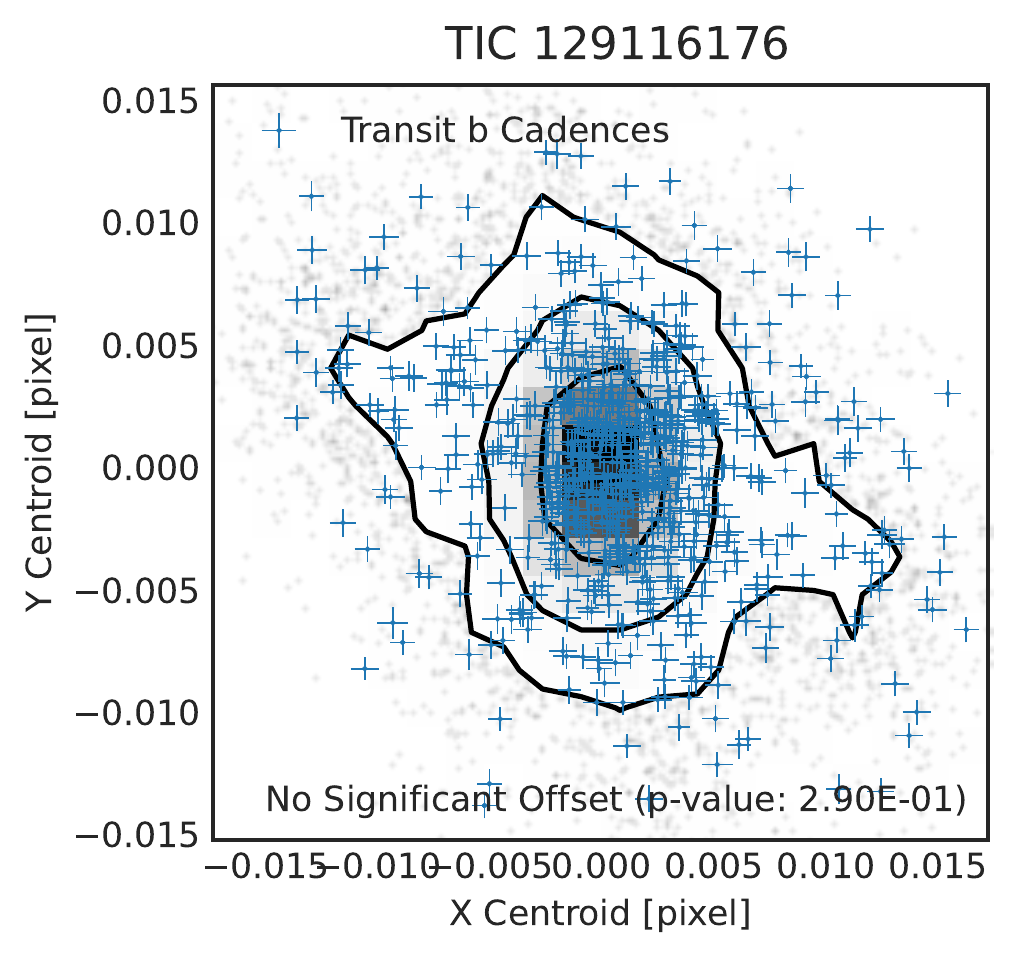}
\endminipage\hfill
\minipage{0.5\textwidth}
  \includegraphics[width=\linewidth]{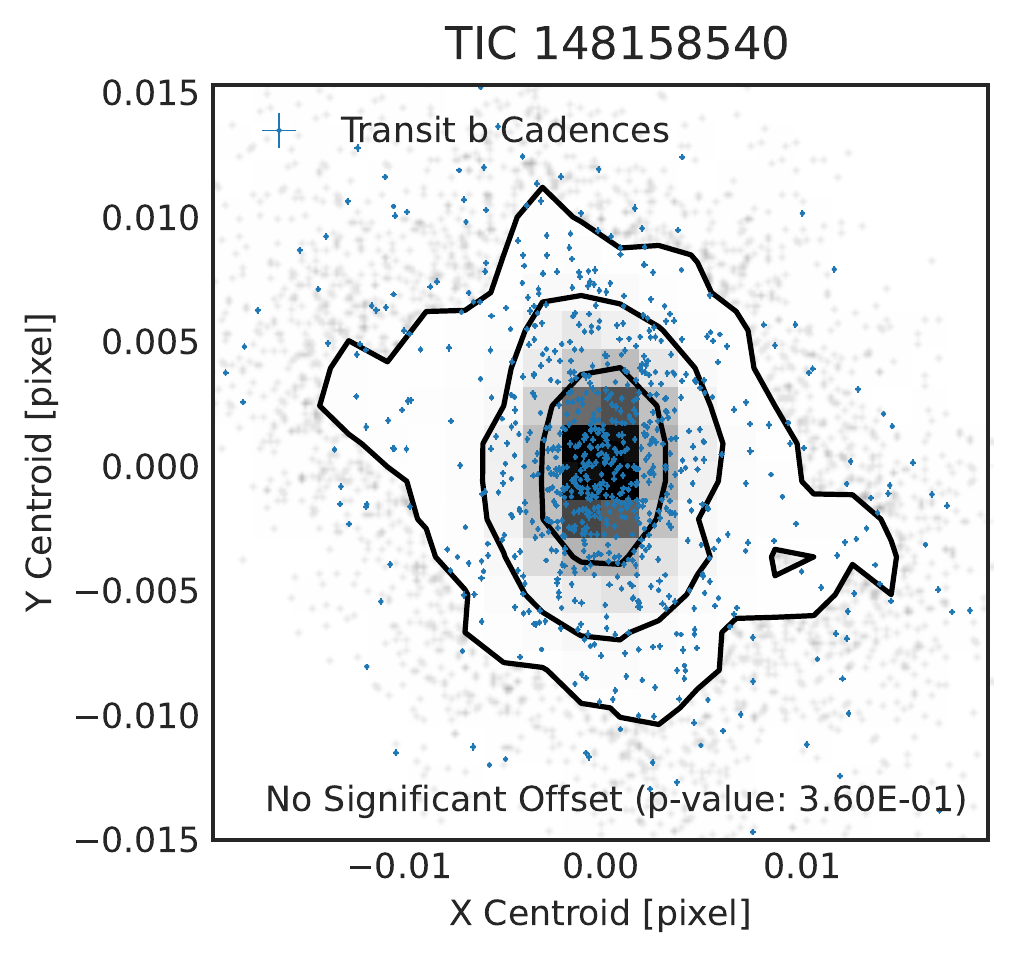}
\endminipage
\caption{Centroid tests for POIs/Potential Binaries in UCL}
\label{fig:centroid_3}
\end{figure}

\clearpage

\section{TESS Input Catalog (TIC) IDs of Cluster Members Searched in this work}\label{sec:tics}
\subsection{Tucana Horologium Association (THA)}
290496318, 266997586, 355766445, 425934574, 281461138, 231867117, 231907973, 231910539, 231914259, 271503441, 234495456, 394210384, 234506344, 234506911, 237880881, 237883772, 237885807, 237910557, 232073492, 234333175, 234338669, 238876720, 274017928, 220523369, 394355443, 238194430, 238196350, 25118964, 33864387, 33881250, 294754396, 319289907, 179315999, 149351776, 141423634, 270623499, 350712873, 150068381, 280865190, 410421717, 79403675, 259845346, 316854083, 406277015, 139984208, 238815210, 327667965, 71274771, 403237836, 234284556, 317016534, 273416539, 197960398, 261560580, 293507177, 229807000, 229807051, 410213494, 410214984, 410214986, 350215424, 325220989, 234290432, 201226029, 325525541, 650381163, 2054867096, 120462356, 12509218, 116157661, 156002545, 117874958, 117874959, 281670243, 206592394, 41862041, 158595208, 206623351, 230982053, 231005004, 229142539, 231020924, 231020638, 219993473, 219998026, 231064749, 231064750, 201753428, 201795667, 201861769, 166787846, 166852312, 166852313, 201898220, 201919099, 201938513, 207138379, 339607562, 381949410, 175298549, 149838027, 224276435, 40059734, 67630561, 140460192, 266746662, 77111651, 152970890, 200790343, 200790342, 401913214, 79566479, 219229275, 219257563, 259542669, 220437346, 381003375, 318785502, 318841562, 141306513, 165184400, 49906916, 176852917, 160008866, 308305105, 93122097, 144653213, 321100028, 44793998, 44794000, 152505948, 170673573, 77339881, 259573397, 161576005, 235056185, 685945639, 632601393, 70596114, 178857861, 117748478, 178969585, 178969584, 406498177, 188969241, 188969240, 77959676, 220622233, 738102496, 397231463, 317325489, 369862028, 451535889, 276967501, 401820403, 401820823, 411956826, 339397736, 466265409, 219953459, 360912453, 265332355, 372106647

\subsection{IC~2602}
269788982, 270294307, 373886624, 374507880, 374434502, 376150973, 376248958, 376731139, 377487978, 463722209, 463914418, 280562557, 281404104, 295146126, 296056755, 303021887, 303938254, 315312718, 315556844, 316108752, 316657304, 338027557, 338153421, 338155185, 344331260, 344500353, 344748662, 371442380, 372091372, 372598349, 372810731, 374748039, 375211992, 375586505, 375590099, 376026697, 376346464, 376802322, 376802477, 377369998, 377370859, 377954373, 378030265, 378229702, 378229708, 378412167, 378421253, 378503608, 378611172, 378968986, 378971350, 389925947, 389927730, 389987955, 389989102, 389989103, 389992758, 389994091, 390287165, 390297263, 390297906, 390356253, 390358373, 390358936, 390359665, 390359863, 390362683, 390362687, 390364291, 390365611, 390366165, 390366768, 390369742, 390370044, 390440616, 390442052, 390442224, 390442286, 390442621, 390443919, 390444472, 390446082, 390756608, 390757974, 390759555, 390759810, 390759889, 390760156, 390760423, 390840249, 390840539, 390841015, 390842457, 390847186, 390849159, 390850870, 390852244, 390853511, 390853705, 390953365, 390953417, 390954082, 390954865, 390955053, 390965584, 390966887, 390967697, 391166737, 391169363, 391247092, 391248084, 391251375, 391252554, 391333587, 391333777, 391334152, 391336297, 391336489, 391341577, 391345891, 391420878, 391421013, 391421894, 391422697, 391423247, 391423941, 391426350, 391426355, 391426724, 391742621, 391742931, 391798710, 391800466, 397179567, 397256049, 397451050, 397607720, 397608769, 397735761, 397781652, 397786354, 398002694, 398003321, 398007252, 398007510, 398009616, 398497171, 398859114, 398913671, 398919018, 398919062, 400066956, 400326671, 401066440, 401885249, 402190411, 402197704, 453950110, 460019760, 460042206, 460042668, 460044249, 460052791, 460053781, 460071039, 460169886, 460179837, 460205515, 460205581, 460205610, 460205613, 460209001, 460228276, 460230188, 460235527, 460297171, 460335450, 460336403, 460337200, 460338822, 460364093, 460366104, 460451570, 460473652, 460474388, 460493086, 460496906, 460500756, 460500920, 460501099, 460502873, 460526824, 460527147, 460613215, 460623602, 460625223, 460648865, 460649476, 460651700, 460651946, 460652126, 460765494, 460771098, 460772349, 460774571, 460774855, 460793411, 460793464, 460795626, 460796730, 460797477, 460797491, 460803319, 460823191, 460823505, 460825967, 460826876, 460827760, 460835832, 460910972, 460950389, 460950671, 460951259, 460953778, 460957968, 460958645, 460982124, 460982877, 460985868, 461080220, 461117076, 461120837, 461121683, 461123552, 461124617, 461125055, 461125245, 461125565, 461129500, 464897777, 464898410, 464900625, 464901555, 464902649, 464903262, 464905819, 464907721, 464907793, 464925976, 464925998, 464928326, 464928360, 464930364, 464934793, 464935013, 464935253, 464935435, 464936128, 464936356, 464936632, 464937455, 464937480, 464938239, 464969661, 465051041, 465069629, 465071629, 465075473, 465100077, 465104773, 465107223, 465107251, 465107412, 465107540, 465107704, 465109422, 465131523, 465132340, 465250214, 465261021, 465263287, 465264828, 465268721, 465268826, 465271210, 465272595, 465273919, 465275413, 465296988, 465297405, 465302289, 465306494, 465451355, 465456282, 465458552, 465459529, 465461596, 465462490, 465466499, 465466759, 465466991, 465470619, 465470793, 465472920, 465512275, 465593684, 465621749, 465661670, 465663582, 465801421, 465801835, 465803266, 465823847, 465828067, 465828262, 465831049, 465865592, 465902398, 465998642, 466002083, 466003005, 466010610, 466030199, 466032289, 466034300, 466035035, 466181275, 466183196, 466186713, 466194095, 466208965, 466209622, 466215841, 466218647, 466368276, 466370148, 466422383, 466568973, 466573354, 466576983, 466577068, 466579089, 466593335, 466617413, 466617428, 466630156, 466705111, 466736533, 466794262, 466831513, 466953672, 466982490, 467136716, 467138305, 467162035, 467170455, 467232072, 467342418, 467349410, 467351743, 467557909, 467739360, 467953288, 467976475, 467980197, 468093585, 468172174, 468205592, 844245965, 844522717, 847701323, 847769574, 848933728, 849260294, 850363872, 850582149, 850582150, 911292786, 911713906, 911870847, 911870867, 911872987, 911894290, 911920053, 911976941, 912047875, 912047923, 912058973, 913701079, 913996012, 914038861, 915086533, 923545851, 374517370, 374930544, 375679671, 376420337

\subsection{Upper Centaurus Lupus (UCL)}
410464477, 231813631, 436248798, 157209108, 284422031, 60024614, 53369126, 140295287, 45688827, 280207697, 446764692, 188084200, 48781344, 49046625, 448474635, 178659732, 29982064, 333734230, 437266963, 847323329, 872689258, 73650567, 73710370, 103625477, 124286087, 124512998, 124515557, 134976179, 142391542, 144609763, 144678969, 144921815, 152484504, 152484511, 162093865, 162208214, 178759525, 181229149, 248078069, 301219220, 323245745, 333832656, 390009811, 451452509, 452419080, 461125552, 911963409, 940640910, 998337743, 998337745, 388237166, 454856551, 273349137, 361818250, 258588387, 442062521, 287670422, 245298317, 339301537, 245430125, 363781484, 275705109, 438586540, 438692318, 243164696, 243237647, 165729961, 165775281, 243321910, 165823257, 243330489, 243344284, 243341756, 101940827, 243389982, 243390044, 243393076, 243396962, 243412695, 243427058, 243430178, 243477650, 243496587, 243511069, 241511379, 241511375, 30067072, 241570607, 241573601, 243618821, 207786920, 166447153, 243642384, 243647020, 241640242, 166577188, 241677723, 276004385, 241685708, 111947267, 307143242, 241739877, 113350106, 113277027, 166843863, 420522238, 359830202, 178931743, 359762838, 112176566, 178936965, 261389834, 261389883, 242119768, 326467176, 326742300, 242149069, 179099805, 179130687, 312420298, 312512588, 179220045, 396137563, 328700272, 179291515, 328811406, 242267549, 330123048, 329299765, 179367009, 179413040, 179413905, 179538968, 329815628, 210806853, 242336470, 242367242, 179623765, 179623744, 323269451, 242391843, 179743858, 339413946, 330987798, 331335557, 339530484, 179878718, 413285274, 331523868, 179905128, 332012383, 242548962, 179964001, 242557720, 332213483, 414138813, 125757999, 242583230, 242583644, 167480388, 242595135, 242612195, 242610390, 167546864, 420582698, 167596713, 332888505, 167670025, 90800382, 127024985, 167704195, 241827885, 241825308, 127089264, 333544395, 127203998, 264137167, 127249954, 127311608, 127246209, 211517703, 127315102, 290932786, 457557304, 241883618, 127664368, 83863971, 127731749, 392788625, 291909020, 158809330, 127794299, 127861722, 127791732, 392838495, 128079374, 158979332, 128119540, 159043143, 159195919, 289425666, 430829746, 369971171, 128453434, 128622430, 379593781, 460844584, 159376787, 159372538, 159421340, 159428617, 159428519, 383492888, 383582909, 159496289, 129116176, 129117325, 129060941, 159537293, 188212197, 188991120, 159752939, 383632304, 159754676, 47752102, 129419398, 47828022, 81003, 47884213, 250091359, 129623324, 366011440, 366011855, 250123846, 160089427, 48064387, 129774131, 129948345, 455309397, 160444470, 160404487, 160444291, 455375884, 455410480, 48364505, 160502580, 160504689, 48437482, 309456404, 164844, 160695823, 307425592, 333651699, 440863421, 276130740, 276219648, 121094784, 121102482, 461663855, 35001854, 440880146, 121240521, 121196226, 334129050, 121250827, 334425144, 334424163, 334417833, 121407551, 75543709, 75540884, 334648804, 75616091, 334835256, 75642561, 75648725, 335146096, 335139519, 75735567, 370476997, 121779963, 121779961, 121726626, 335454137, 121839093, 75869475, 335643748, 121982262, 121981993, 75873633, 75931110, 121983359, 121979152, 75933042, 75970186, 75971901, 61012630, 61079477, 160417350, 160417246, 160549804, 76050403, 122083452, 122162363, 76057224, 160633268, 160701354, 160704414, 122244909, 160632970, 160703925, 122250596, 160703933, 121850125, 122288890, 122397232, 122394390, 122397233, 76322064, 76397021, 276293815, 371122327, 276300910, 76435066, 276452267, 276491524, 276503464, 76580025, 141662588, 76587359, 307526685, 307533179, 61530691, 76652075, 142513032, 143414073, 76706808, 76706801, 272341056, 148087359, 148158374, 148158540, 272397344, 272395377, 148158169, 48630780, 272396062, 148261121, 272456799, 272462967, 460329848, 460323913, 225797367, 148480904, 148494823, 225957292, 148495018, 89026796, 148772375, 54172975, 54175032, 54175033, 148904033, 292265209, 292364079, 148904850, 148904891, 54245747, 54449487, 185053184, 54667962, 54810190, 54811378, 1037289516, 976426321, 128116539, 1052698599, 1174609023, 127871423, 42974961, 133010063, 132241629, 149640056, 141045993, 413145172, 461728725, 413693742, 413693728, 414077635, 152377452, 153079294, 149437620, 153781478, 154438290, 155430355, 170986304, 289877522, 172608950, 172608996, 189421351, 172615355, 173560788, 173560502, 173871956, 254322223, 254365686, 254406143, 254387071, 190372855, 176463845, 254451455, 98761335, 98909539, 177342508, 177336303, 177631209, 254597659, 99039745, 99039926, 178201756, 99041233, 99039843, 266271836, 99039842, 99046173, 99046104, 362351038, 254628801, 99205053, 178456568, 269155561, 269155560, 179051609, 179051461, 270107026, 99324522, 179622269, 272977820, 254740868, 254771390, 442611273, 295901750, 295905052, 295899674, 254844960, 295911223, 277674241, 296039638, 442642949, 254930869, 254901458, 296483317, 296700769, 255092900, 58413338, 442813719, 301036163, 301087756, 301048970, 255134616, 442828160, 58583881, 364200069, 364200072, 364200513, 255263576, 255236579, 58751944, 255246336, 292566342, 386943436, 58752825, 58749494, 386980599, 58832313, 294993587, 281164246, 281164807, 279883855, 255403319, 279819212, 279821618, 255407474, 59017926, 59095543, 59095516, 59183599, 59232730, 255464386, 59334299, 93457683, 93456758, 275018323, 255475779, 59403154, 59336031, 59504074, 59500324, 93610654, 93700013, 93779101, 93778206, 69283667, 69377602, 69178294, 256902444, 69420071, 93932740, 214870481, 93985722, 69514028, 69609942, 94159843, 161786522, 161873315, 70041394, 374542527, 374648559, 94491270, 374823554, 374812449, 374719920, 215392772, 374732772, 94731904, 94683443, 457995946, 457993297, 67978974, 382643945, 163141448, 382799779, 382794993, 382870216, 382800509, 382800518, 462044648, 462077825, 462083149, 462147302, 462148256, 462146157, 462147898, 462146834, 68256179, 164029492, 95468062, 382549695, 303188383, 164707310, 303188775, 68416596, 164707361, 318027415, 318030913, 317906231, 318035342, 68565222, 3084844, 410857373, 364368528, 364523302, 96552404, 318141346, 318141352, 96631952, 210541258, 210577071, 221355243, 3926262, 4061225, 222835418, 210908741, 210908710, 222940415, 392328254, 350869718, 210975114, 2743109, 211086449, 224004929, 351521484, 211143017, 211237627, 211237621, 225109252, 352162554, 211459804, 323187268, 211609345, 211681133, 291866634, 280474618, 292925229, 280828479, 280822724, 212201942, 382951562, 205461071, 83687552, 310053760, 458058639, 94972416, 233981889, 397712527, 234869424, 78981730, 79358659, 245927009, 245934292, 337243296, 338104678, 338105219, 246664608, 338495232, 34404183, 339989392, 247348620, 343976519, 212461524, 377523286, 378094374, 381441805, 381441323, 44207349, 193169513, 194744695, 160689461, 17509396, 1232119536, 1233589158, 1336147896, 1343250594, 1160500678, 1247277024, 504761559, 1170473308, 1172055696, 1255305471, 446547327, 469811127, 148910632, 97044158, 285117481, 393808105

\subsection{Ursa Major (UMa)}
11895653, 224288693, 224292240, 316331312, 224305606, 99381773, 229534764, 1001230963, 157272202, 159190005, 159191044, 141821609, 150387644, 142277151, 159189482, 416538839

\subsection{Pisces-Eridani (PiEri)}
80363380, 41858251, 262531383, 12787754, 399570739, 240720758, 4579839, 188627184, 311085703, 8938429, 66437682, 66471572, 89526411, 228383334, 89547213, 89547209, 114749459, 37763129, 212963433, 382306185, 28077586, 28092085, 248349830, 67643537, 248384545, 248952247, 70490059, 257434109, 29848163, 404467056, 164777021, 33625968, 423750380, 423750384, 257483561, 257484419, 72729895, 139526229, 120888774, 175491080, 144110354, 630138333, 332714671, 332740776, 270432631, 1028816, 332825272, 332862003, 332864335, 420000810, 1122408, 251064437, 251081955, 64008790, 251088243, 4741056, 65346691, 425095257, 35745429, 35841788, 204613238, 204613241, 141287580, 98855268, 365332374, 23193695, 9805084, 274037216, 441515106, 92250588, 279068672, 55893164, 9967295, 423442532, 12546422, 382326499, 12631414, 121006547, 166741952, 325707772, 156460685, 165219666, 279403260, 44647574, 461713170, 178286446, 257605131, 152590946, 170674625, 651901104, 457161840, 38613715, 55406666, 55441489, 34086013, 34147414, 9161614, 34238114, 34267639, 269724428, 332661361, 250133028, 37816550, 37860207, 248613540, 37903382, 37948770, 71089681, 71091056, 248655215, 38049622, 38049612, 38128400, 55935823, 117735557, 117745508, 56000236, 56127663, 146804027, 167485031, 246896530, 316915529, 152439649, 169396430, 169451121, 169529956, 231303250, 188968885, 188987821, 24607967, 442926120, 442926121, 24719594, 442965496, 143353370, 408233707, 437759888, 385141581

\clearpage
\end{document}